\documentclass[lettersize,journal]{IEEEtran}
\usepackage{amsmath,amsfonts}
\usepackage{algorithmic}
\usepackage{algorithm}
\usepackage{array}
\usepackage[caption=false,font=normalsize,labelfont=sf,textfont=sf]{subfig}
\usepackage{textcomp}
\usepackage{stfloats}
\usepackage{url}
\usepackage{verbatim}
\usepackage{graphicx}
\usepackage{cite}
\usepackage{bm}
\usepackage{multirow}
\usepackage{algorithm}
\usepackage{algorithmic}
\usepackage{fancyhdr}

\usepackage{soul, color, xcolor}
\soulregister{\cite}7 
\soulregister{\citep}7 
\soulregister{\citet}7 
\soulregister{\ref}7 
\soulregister{\pageref}7 

\graphicspath{{figures/}}
\begin{document}

\title{Sparse Bayesian Learning-Based Hierarchical Construction for 3D Radio Environment Maps Incorporating Channel Shadowing}

\author{Jie Wang, Qiuming Zhu,~\IEEEmembership{Senior Member,~IEEE,} Zhipeng Lin,~\IEEEmembership{Member,~IEEE,} Junting Chen,~\IEEEmembership{Member,~IEEE,} Guoru Ding, Qihui Wu,~\IEEEmembership{Senior Member,~IEEE,} Guochen Gu, Qianhao Gao
  \thanks{This work was supported in part by the National Key Scientific Instrument and Equipment Development Project under Grant No. 61827801, in part by the National Natural Science Foundation of China under Grant No. 62271250, in part by the Key Technologies R\&D Program of Jiangsu (Prospective and Key Technologies for Industry) under Grants BE2022067, BE2022067-2 and BE2022067-4, and in part by Practice Innovation Program of Jiangsu Province under Grant KYCX23\_0381. (\emph{Corresponding author: Qiuming Zhu})}
  \thanks{J. Wang, Q. Zhu, Z. Lin, Q. Wu, G. Gu and Q. Gao are with The Key Laboratory of Dynamic Cognitive System of Electromagnetic Spectrum Space, College of Electronic and Information Engineering, Nanjing University of Aeronautics and Astronautics, Nanjing 211106, China  (e-mail: {bx2104906wangjie, zhuqiuming, linlzp}@nuaa.edu.cn, wuqihui2014@sina.com, {guguochen, qianhaogao}@nuaa.edu.cn).}
  \thanks{J. Chen is with The School of Science and Engineering, and the Future Network of Intelligence Institute (FNii), The Chinese University of Hong Kong, Shenzhen, Guangdong 518172, China (e-mail: juntingc@cuhk.edu.cn).}
    \thanks{G. Ding is with the College of Communications Engineering, Army Engineering University, Nanjing 210007, China (e-mail: guoru\_ding@yeah.net).}
}

\markboth{}%
{Wang \MakeLowercase{\textit{et al.}}: Sparse Bayesian Learning-Based Hierarchical Construction}

\IEEEpubid{0000--0000/00\$00.00~\copyright~2021 IEEE}

\fancyhf{}
\fancyhead[LE]{\leftmark} 
\fancyhead[RO]{\rightmark} 
\fancyfoot[LE,RO]{\thepage} 

\maketitle

\begin{abstract}
The radio environment map (REM) visually displays the spectrum information over the geographical map and plays a significant role in monitoring, management, and security of spectrum resources.
In this paper, we present an efficient 3D REM construction scheme based on the sparse Bayesian learning (SBL), which aims to recover the accurate REM with limited and optimized sampling data. 
In order to reduce the number of sampling sensors, an efficient sparse sampling method for unknown scenarios is proposed. For the given construction accuracy and the priority of each location, the quantity and sampling locations can be jointly optimized. 
With the sparse sampled data, by mining the sparsity of the spectrum situation and channel propagation characteristics, a SBL-based spectrum data hierarchical recovery algorithm is developed to estimate the missing data of unsampled locations. 
Finally, the simulated 3D REM data in the campus scenario are used to verify the proposed methods as well as to compare with the state-of-the-art. We also analyze the recovery performance and the impact of different parameters on the constructed REMs. Numerical results demonstrate that the proposed scheme can ensure the construction accuracy and improve the computational efficiency under the low sampling rate.

\end{abstract}


\section{Introduction}
\subsection{Background and Motivation}
\IEEEPARstart{W}{ith} the rapid development of Radio Frequency (RF) communication technologies, various electronic devices, i.e., radio, radar, navigation, and so on, have formed a dynamic, complex, and multidimensional electromagnetic environment \cite{Ding18CM, Ahmad15CST, Wang11JSTSP, Liu23TWC}. Nevertheless, a large percentage of assigned spectrum remains underutilized as reported by the Federal Communications Commission (FCC) \cite{Kolodzy02FCC}. Consequently, the dynamic spectrum access technology, i.e., cognitive radio (CR), was introduced to improve communication efficiency based on the awareness of spectrum situation \cite{Ahmad15CST, Huang23SJ, Huang24space}. The Defense Advanced Research Projects Agency (DARPA) RadioMap program further proposed to add the spectrum awareness results on a geographical map, namely radio environment map (REM) or spectrum cartography (SC) \cite{Yilmaz13CM}. In a word, the REM can visualize specific spectrum information, i.e., received signal strength (RSS), channel gain, power spectrum density, and radiation {location, in a region of interest (ROI) at different dimensions and scales from limited observations. It is essential in dynamic spectrum access, abnormal spectrum monitoring, spectrum sharing, radiation localization, interference coordination, etc\cite{He23TWC, Pan22TWC}.  
\IEEEpubidadjcol

Majority of SC methods focused on the two-dimension (2D) REM construction, which usually treated it as solving a matrix completion problem or image reconstruction task based on sufficient sampled data\cite{Romero22SPM}. However, with the development of space-air-ground integrated networks, the spectrum situation has been extended from 2D space to three-dimension (3D) space\cite{Ding16JSAC,Shen22TWC}. For example, the aerial platforms, i.e., unmanned aerial vehicle (UAV), airship, and hot air balloon, have becoming an important part of radiation sources\cite{Shi23TC}. Besides, the increased spatial freedom and enhanced spatial isolation between Primary Users (PUs) and Secondary Users (SUs) enable the exploration of spectrum access opportunities in 3D space, which were not present in the 2D plane. The 3D REM facilitates the implementation of spectrum reuse schemes, dynamic spectrum access, and the prediction of optimal network connections. An illustrative example is a user handover algorithm proposed in\cite{chen15access}, which leverages the 3D REM to predict the best network connection. Therefore, it is essential to study the 3D REM construction in the unknown spatial heterogeneous environment.

Due to the limitation of sensor number and sampling time in the wide space, it is a big challenge to achieve accurate reconstruction with sparse and noisy data \cite{Bi19WC}. To address this problem, compressed sensing (CS) technology has been extensively adopted since it requires less data than the Nyquist criterion\cite{Alfonso18Sensors, Bazerque10TSP}. By mining the underlying sparsity of spectrum situation, the CS constructs the REM by decomposing the sampling data into the linear superposition of sensing matrix and sparse signal. Nevertheless, the sensing matrix exhibits high spatial correlation and the recovery performance could be greatly deteriorated when dealing with the noisy data. Sparse Bayesian learning (SBL) can recover the exact sparse signal under the high correlated sensing matrix and has good performance of anti-noise\cite{Tipping01JMLR}. 

Moreover, the measurements are spatially correlated, and thus the sampling locations have a great impact on the accuracy and efficiency of construction. By selecting specific sampling locations, it is possible to recover the 3D REM with less data requirements. Besides, a poor propagation model may also lead to significant construction errors\cite{Pang21WC, Chen19TWC}. In the specific scenarios, i.e., dense urban areas with numerous buildings, it is difficult to build an accurate channel model including the factors of multi­building, shadowing, and antennas pattern. 

To sum up, the available sampling data in 3D space is sparse. The methods used for 2D REM are not suitable for 3D REM construction. Besides, the propagation model of existing 2D (or ground) REM cannot satisfy the dense 3D RF environment and may cause serious construction errors. According to the above observations, we are motivated to exploit the inherent sparsity of spectrum situation, the radio propagation characteristics, and sampling position optimization for constructing accurate 3D REMs in unknown environments. 
 
\subsection{Related Work}
Existing REM construction methods can be mainly divided into data-driven and model-driven ones\cite{ DALLANESE11TVT, Shrestha22TSP, HAN20Sensors, YILMAZ15WCMC, PESKO15EURASIP, Shen22TWC, Romero22TWC, DEMO22, zhao2022temporal, Chen22TSP, Naranjo14WCNC, Sato17TCCN, Fuchs17TAP, TANG16Access, Huang15TVT, wang23sparse}. The former typically employs the spatial interpolation and machine learning (ML) techniques to recover the missing data at the unsampled positions by the available sampling data. In\cite{Naranjo14WCNC}, the authors compared four typical spatial interpolation techniques, i.e., Nearest Neighbour (NN), inverse distance weighting (IDW), triangular irregular network, and Kriging algorithm\cite{Sato17TCCN}. The authors also mentioned that Kriging was the most accurate spatial interpolation method while IDW was robust. In \cite{Fuchs17TAP}, a method is proposed to construct the REMs with thin plate spline interpolation when no information on the radiation source is available. The authors in\cite{TANG16Access} proposed a tensor completion method to recover the incomplete spectrum data in both the spatial and temporal domains. 

Recently, ML techniques recast data complementation as a learning-based optimization problem and have also been applied in REM construction. For example, a generative adversarial network-based method was proposed in\cite{HAN20Sensors} to recover the missing spectrum data from the simulated or previously collected training data. The authors in\cite{Shrestha22TSP} utilized deep neural networks (DNNs) to learn the intricate underlying structure from the sampled data. In\cite{Shrestha23TWC}, the authors further proposed a DNN-based active sensing method with UAVs. However, these methods require a large amount of sampling data, in order to achieve satisfactory performance. Besides, the construction performance are usually sensitive to the sampling position distribution and measurement error\cite{Bi19WC}.

The model-driven methods combine the sampling data with active transmitter information and channel propagation models to construct REMs, which are less sensitive to the measurement error as well as the number of sampling data. For example, the authors in\cite{YILMAZ15WCMC} firstly estimated the transmitter information by the sampled data and then completed the missing data with the help of propagation model. CS is a typical model-driven construction method, which can track the dynamic network topology and construct the REM with sparse sampled data. In\cite{Alfonso18Sensors}, CS multispectral cartography was proposed for spectrum sensing. The least absolute shrinkage and selection operator was used in\cite{Bazerque10TSP} to construct the REM with random sampling data. In\cite{Shen22TWC}, the authors proposed a compressed REM mapping method based on the improved orthogonal matching pursuit algorithm. However, both of two are easy to be effected by the noisy measurement and high-correlation sensing matrix. Besides, they only focused on the REM construction under line of sight (LOS) conditions without considering the effect of a realistic environment on the channel propagation.

The sparse Bayesian learning (SBL) can recover the sparse signal under the high correlated sensing matrix. The authors in\cite{Huang15TVT} proposed an SBL-based 2D REM construction method. The simple path loss model was considered and random sampling was performed. Besides, it assumed that each spatial location was only affected by single transmitter, which is impractical in the actual scenarios. In\cite{wang23sparse}, a scenario-dependent SBL method was proposed to construct 3D REM. It considered the realistic propagation channel by using the ray tracing (RT) technology on the known scenario. The authors also developed a sampling location optimization algorithm. In\cite{He18INFOCOM}, a Bayesian compressive crowdsensing framework was applied to address the crowdsourced 2D REM construction. 

In a nutshell, most of existing studies only focus on the 2D REM construction and the measurement positions or matrixes are either random or fixed without considering the sampling optimization. Furthermore, they are generally based on the free-space channel model or RT-based model. This is only applicable to flat suburban or other known scenarios. Especially, the channel fading characteristic, i.e., shadow fading, is not considered under unknown scenarios to the best of our knowledge.

\subsection{Contributions}
To fill these gaps, this paper proposes a novel SBL-based 3D REM hierarchical construction scheme. It consists of the sampling location optimization and the spectrum data recovery with consideration of shadow fading. The main novelties and contributions of this paper are summarized as follows.
\begin{itemize}	
	\item An efficient SBL-based hierarchical construction model for 3D REM in the unknown spatial heterogeneous environment is proposed. Different with previous methods, the proposed method considers the factors of sampling location optimization and channel propagation characteristic including shadow fading, which can improve the construction performance with the limitation of sampling data and time.
	
	\item A worst case SBL variance-based measurement matrix optimization method is proposed. By introducing the minimum SBL variance indicator and solving a sparse optimization problem, the optimized sampling locations and quantity are obtained under the given threshold of sparse signal recovery accuracy. The proposed method outperforms other sampling methods and improves the sampling efficiency while ensuring the recovery accuracy.
	
	\item A spectrum data hierarchical recovery algorithm under unknown scenarios is developed based on the Bayesian theory. By mining the sparsity characteristics in the 3D spectrum space, the sparse signal is firstly recovered based on the SBL. In order to obtain the whole 3D REM, channel dictionary is then optimized by considering the channel characteristic and shadowing based on Gaussian process. The proposed algorithm exhibits excellent recovery performance against the traditional algorithms especially under unknown scenarios. 
\end{itemize}

The rest of this paper is organized as follows. Section II gives the proposed 3D REM construction model. In Section III, the details of proposed construction scheme are given and demonstrated. Then, Section IV presents the simulation and comparison results and Section V gives some conclusions.

\begin{figure*}[!t]
	\centering
	\includegraphics[width=5.5in]{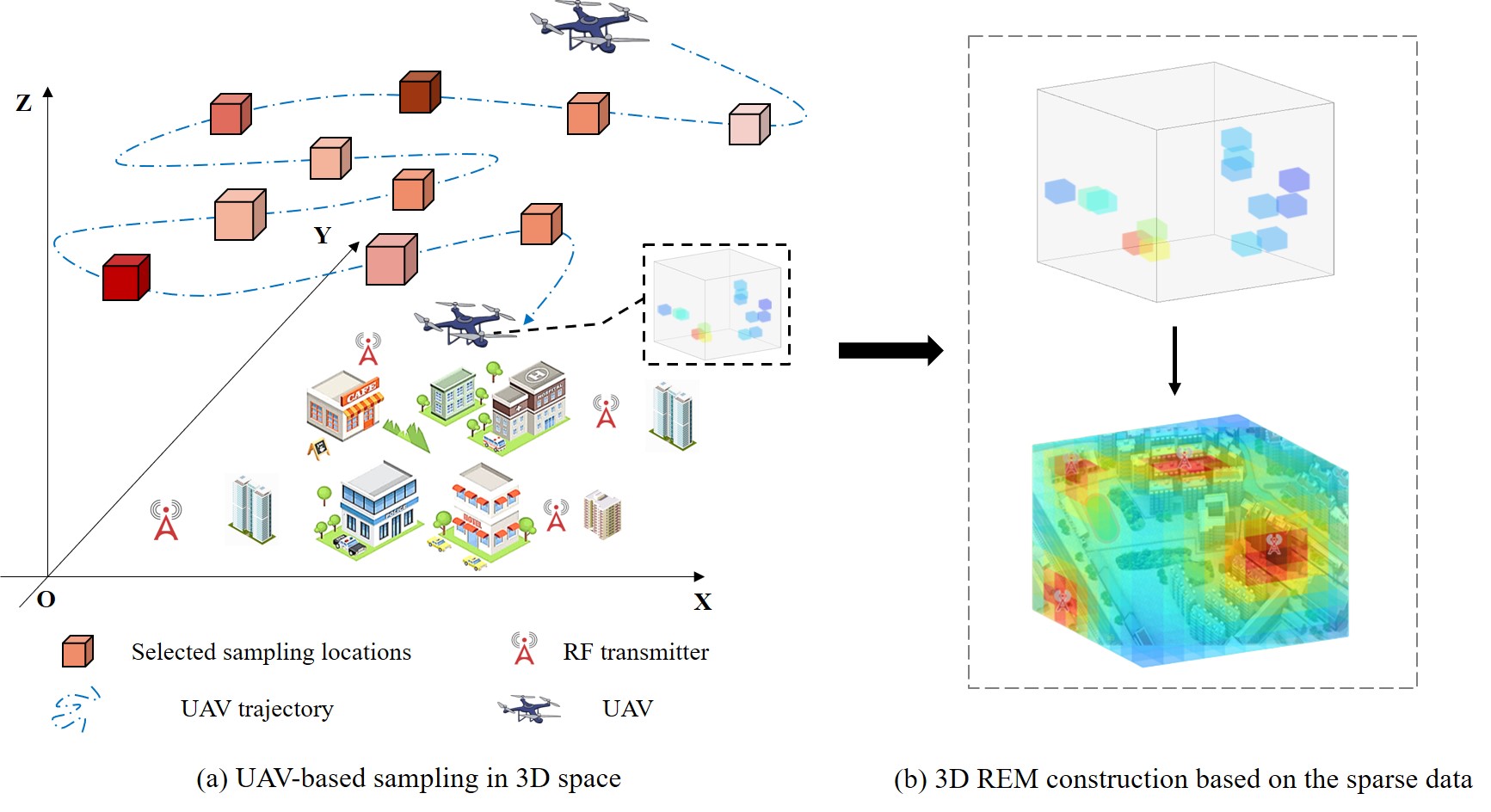}
	\caption{Overview of UAV-based 3D REM sampling and construction.}
	\label{fig1}
\end{figure*}

\section{PRELIMINARIES}
\label{SYSTEM}
\subsection{Proposed CS-Based 3D REM Construction}
The 3D region of interest (ROI) is firstly discretized into several small cubes. Thus, it can be described by a 3D spectrum tensor $\bm{\chi}  \in {\Re ^{{N_x} \times {N_y} \times {N_z}}}$, where ${N_x}$, ${N_y}$, and ${N_z}$ indicate the grid number along $x$, $y$, and $z$ dimensions, respectively. Each element or cube is colored according to its RSS value. Technically, 3D REM construction aims to recover all RSS values, i.e., $N = {N_x} \times {N_y} \times {N_z}$ cubes, based on the known ones of sampled cubes. The locations of the all cubes are demoted as $\left\{ {{{\mathbf{\bm{\nu} }}_n}} \right\}_{n = 1}^N$, where ${{\mathbf{\bm{\nu} }}_n} = \left( {x_n^v,y_n^v,z_n^v} \right)$. Due to the low cost, mobility, and flexibility movement, an UAV equipped with spectrum monitoring device can be used to sample RSS values in the selected cubes of ROI as shown in Fig. \ref{fig1}. Here, RSS signifies the power strength of the spectrum. Note that this paper only focuses on how to select optimal sampling points, without considering the trajectory planning of UAVs. In practical applications, after all optimized sampling points are determined, some optimization algorithms such as shortest path planning can be further utilized to obtain the trajectory of UAV for data acquisition.

Let us set a sparse signal vector $\bm{\omega}={\left[ {{\omega_1},{\omega_2}, \ldots ,{\omega_n}, \ldots ,{\omega_N}} \right]^{\text{T}}} \in {\mathbb{R}^{N \times 1}}$ to denote the unknown transmitting information at every cube as
\begin{equation}
{\omega _n} = \begin{cases}
{P_n^t,}&{\text{if there is a RF transmitter in the }}n{\text{th cube,}} \\ 
{0,}&{\text{else,}} 
\end{cases}
\label{eq1}
\end{equation}
where $P_n^t$ is the transmitting power of RF transmitter if there is a RF transmitter in the $n$th cube. It is assumed that there are $K$ stationary RF transmitters in the 3D ROI denoted by $\left\{ {{{\mathbf{u}}_k}} \right\}_{k = 1}^K$, where ${{\mathbf{u}}_k} = \left( {x_k^u,y_k^u,z_k^u} \right)$ is the location of $k$th RF transmitter. Compared with the cube number in the 3D ROI, the RF transmitter number is much smaller, i.e., $K\left( {K \ll N} \right)$, which means that \bm{$\omega$} has stronger sparsity than in 2D ROI. Therefore, \bm{$\omega$} is a $K$-sparse signal vector with $\left\| \bm{\omega}  \right\|_0^{} = K$.

In the realistic 3D environments, the RSS in each cube is affected by RF transmitter positions and distribution, transmitting power, and channel propagation characteristic. Especially, the channel characteristic mainly considers the channel gain, which includes path loss, shadow fading, small-scale fading, and noise \cite{Hua23TC}. Path loss is a deterministic component determined by frequency and distance between the transmitter and receiver. Small-scale fading and noise can be averaged out with long-term observation, while shadow fading is related with specific scenarios\cite{ZanellaCST_16}. What’s more, there are more complex shadow effects in the 3D scenarios, including the occlusion of signals by buildings, terrain, etc., and the propagation model is essential to effectively incorporate these shadow effects to reflect the propagation of the signal more realistically.

Based on the above analysis, the spectrum tensor $\chi $ is vectorized into ${\mathbf{x}} \in {\mathbb{R}^{N \times 1}}$, as
\begin{equation}
{\mathbf{x}} = \bm{\xi} _{}^v \circ \left( {\bm{\varphi} \bm{\omega} } \right),
\label{eq2}
\end{equation}
where $\bm{\varphi}  \in {\mathbb{R}^{N \times N}}$ is path loss matrix and ${\varphi _{i,j}}$ is defined as the path loss from the $i\text{th}$ cube to the $j\text{th}$ cube. $\bm{\xi} _{}^v \in {\mathbb{R}^{N \times 1}}$ denote the shadow fading components of all cubes, and $ \circ $ denote the element-wise (Hadamard) product. We define the term $\bm{\xi} _{}^v \circ \bm{\varphi} $ as the channel propagation dictionary matrix.

Suppose that we select $M$ cubes as samples, denoted by $\left\{ {{\bm{s}_m}} \right\}_{m = 1}^M$, from all $N$ cubes and ${\bm{s}_m} = \left( {x_m^s,y_m^s,z_m^s} \right)$ is the location of $m\text{th}$ cube. Thus, the sampling rate is $r = M/N$. The sampled RSSs can be expressed by a vector $\bm{\tilde t} \in {\mathbb{R}^{M \times 1}}$. All sampled locations can be represented by a measurement matrix $\bm{\psi} \in {\mathbb{R}^{M \times N}}$ as
\begin{equation}
{\psi _{i,j}} = \left\{ {\begin{array}{*{20}{l}}
	{1,{\text{ if the }}i\text{th}{\text{ sample is at the }}j\text{th}{\text{ cube,}}} \\ 
	{0,{\text{ else,}}} 
	\end{array}} \right.
\label{eq3}
\end{equation}
where each row of $\bm{\psi}$ has a nonzero element denoting the sampling location. Note that the sampling strategy for 2D space is relatively simple, such as uniform sampling or random sampling. As the signal varies in 3D space and the spatial correlation between sampling locations is stronger, the measurement matrix $\bm{\psi}$ should be well designed to ensure efficient sampling of the 3D ROI.

The averaged RSS vector $\bm{t}$ can be expressed as
\begin{equation}
\begin{aligned}
\bm{t} &= \bm{\psi} \left( {\bm{\xi} _{}^v \circ \bm{\varphi} \bm{\omega} } \right) + \bm{\varepsilon}  = \bm{\psi} \bm{\xi} _{}^v \circ \left( {\bm{\psi} \bm{\varphi} \bm{\omega} } \right) + \bm{\varepsilon} \\ 
&= \bm{\xi} _{}^s \circ \left( {{\mathbf{\Phi \bm{\omega} }}} \right) + \bm{\varepsilon}  = {\mathbf{\Phi }}\bm{\omega}  + \bm{\varepsilon} _{}^ * ,
\label{eq4}
\end{aligned}
\end{equation}
with
\begin{equation}
{t_m} = \xi _m^s\sum\limits_{n = 1}^N {{\omega _n}{\Phi _{m,n}}}  + {\varepsilon _m},
\label{eq7}
\end{equation}
\begin{equation}
\begin{array}{*{20}{c}}
{\varepsilon _m^ *  = \left( {\xi _m^s - 1} \right)\sum\limits_{n = 1}^N {\left( {{\Phi _{m,n}}{\omega _n}} \right)}  + {\varepsilon _m},}&{m = 1,2,...,M} 
\end{array},
\label{eq6}
\end{equation}
where $\bm{\varepsilon} \in {\mathbb{R}^{M \times 1}}$ is the measurement noise vector, and ${\mathbf{\Phi }}$ is the sensing matrix. $\xi _m^s$ is the shadow fading value at the sample whose 
logarithmic form follows normal distribution $\mathcal{N}\left( {0,\sigma _{}^2} \right)$. Then, the $\xi _m^s - 1$ can be derived as
\begin{equation}
\xi _m^s - 1 = \exp \left\{ {\frac{{\ln 10}}{{10}}\bar \xi _m^s} \right\} - 1 \sim \frac{{\ln 10}}{{10}}\bar \xi _m^s,
\label{eq66}
\end{equation}
with
\begin{equation}
\bar \xi _m^s = 10{\log _{10}}\xi _m^s.
\label{eq67}
\end{equation}
We assume that the marginal standard deviation  is small enough, then the characteristic of the original multiplicative noise is very similar to that of the additive Gaussian noise \cite{Jiang21SJ}. Accordingly, we approximate $\varepsilon _m^ * $ as a Gaussian distribution with variance $\sigma _0^2$ since $\bar \xi _m^s$ follows $\mathcal{N}\left( {0,\sigma _{}^2} \right)$. 

In this paper, the objective of REM construction is to recovery sparse signal $\bm{\hat \omega} $ and the shadow fading components $\bm{\xi} _{}^v \in {\mathbb{R}^{N \times 1}}$  of all cubes to recovery ${\mathbf{x}}$. Then, the 3D spectrum tensor $\chi $ is obtained by mapping ${\mathbf{x}}$ to 3D space, which can be achieved by MATLAB's matrix transformation function reshape

\subsection{SBL-Based Hierarchical REM Data Recovery}
\label{SSM}
1) SBL-based sparse signal recovery

The matrix $\bm{\varphi} $ denotes the propagation channel characteristic, which is usually modelled by an exponential attenuation model with respect to the distance \cite{Bazerque10TSP}. The element of $\bm{\varphi} $ can be expressed by path loss function $f\left( {{{\mathbf{\bm{\nu} }}_i},{{\mathbf{\bm{\nu} }}_j}} \right)$ as 
\begin{equation}
{\varphi _{i,j}} = f\left( {{{\mathbf{\bm{\nu} }}_i},{{\mathbf{\bm{\nu} }}_j}} \right) = \left\{ {\begin{array}{*{20}{c}}
	{1,}&{{\text{if }}d_{i,j}^{} \le  {d_0},} \\ 
	{\frac{{{G_t}{G_r}c_0^2}}{{4\pi {{\left( {{f_c}} \right)}^2}}}\left( {\frac{{d_0^{}}}{{d_{i,j}^{}}}} \right)_{}^\eta ,}&{{\text{otherwise,}}} 
	\end{array}} \right.
\label{eq9}
\end{equation}
where ${G_t}$ and ${G_r}$ are the antenna gains of transceivers, respectively, ${c_0}$ is the light speed, ${f_c}$ is the carrier frequency, $\eta $ is the path loss exponent, $d_{i,j}^{} = {\left\| {{{\mathbf{\bm{\nu} }}_i} - {{\mathbf{\bm{\nu} }}_j}} \right\|_2}$ is the distance between ${{\mathbf{\bm{\nu} }}_i}$ and ${{\mathbf{\bm{\nu} }}_j}$, and ${d_0}$ is the reference distance. Similarly,  $f\left( {{\bm{s}_m},{{\mathbf{\bm{\nu} }}_n}} \right)$ constructs the sensing matrix ${\mathbf{\bm{\Phi} }}$ between samples and each cube, i.e., ${\Phi _{m,n}} = f\left( {{\bm{s}_m},{{\mathbf{\bm{\nu} }}_n}} \right)$. 

Based on the sparsity characteristic of $\bm{\omega}$, we can recover the sparse signal $\bm{\omega}$ by 
\begin{equation}
\begin{gathered}
\bm{\hat \omega}  = \arg \min {\left\| \bm{\omega}  \right\|_1}, \hfill \\
\begin{array}{*{20}{c}}
{{\text{s}}{\text{.t}}{\text{.}}}&{\bm{t} = \bm{\psi} \bm{\varphi} \bm{\omega}  + \bm{\varepsilon} _{}^ * } 
\end{array}. \hfill \\ 
\end{gathered} 
\label{eq10}
\end{equation}
The SBL approach is adopted in this paper to solve this problem, since it has good performance even with the high correlation of ${\mathbf{\bm{\Phi}}}$\cite{Tipping01JMLR}. 

Firstly, we introduce the probability model in SBL. The sparse regression model (\ref{eq4}) is usually assumed that measurements are corrupted by i.i.d. Gaussian measurement noise $\bm{\varepsilon} _{}^ * $ with unknown variance $\sigma _0^2$, yielding the Gaussian likelihood of $\bm{t}$
\begin{equation}
p\left( {\bm{t}|\bm{\omega} ,\sigma _0^2} \right) = {\left( {2\pi \sigma _0^2} \right)^{ - M/2}}\exp \left\{ { - \frac{{{{\left\| {\bm{t} - {\mathbf{\Phi }}\bm{\omega} } \right\|}^2}}}{{2\sigma _0^2}}} \right\}.
\label{eq11}
\end{equation}
A Gamma distribution is then posed on $\beta \left( {\beta  = \left( {\sigma _0^2} \right)_{}^{ - 1}} \right)$
\begin{equation}
p\left( {\beta ;{c_0},{d_0}} \right) = \Gamma \left( {\beta |{c_0},{d_0}} \right),
\label{eq12}
\end{equation}
with
\begin{equation}
\Gamma \left( {\beta |{c_0},{d_0}} \right) = \Gamma {({c_0})^{ - 1}}{d_0}^{{c_0}}{\beta ^{{c_0} - 1}}{e^{ - {d_0}\beta }},
\label{eq13}
\end{equation}
where $c_0 \geq 0$ and $d_0 \geq 0$ are the shape parameter and the scale parameter, respectively, $\Gamma \left(  \cdot  \right)$ is the Gamma function $\Gamma ({c_0}) = \int_0^\infty  {{t^{{c_0} - 1}}{e^{ - t}}dt}$. 

Then, to induce the sparsity of $\bm{\omega}$, we deploy a sparseness-promoting prior on it, i.e., a two-layer hierarchical sparse prior \cite{Tipping01JMLR}. In the first layer, each element of $\bm{\omega}$ is posed a zero-mean Gaussian prior as
\begin{equation}
p\left( {\bm{\omega} |\bm{\alpha} } \right) = \prod\limits_{i = 0}^N {\mathcal{N}\left( {{\omega _i}|0,\alpha _i^{ - 1}} \right)} ,
\label{eq14}
\end{equation}
where $\bm{\alpha}  = \left[ {{\alpha _1},{\alpha _2}, \ldots ,{\alpha _N}} \right]_{}^{\text{T}}$. Then, a Gamma hyperprior over $\bm{\alpha}$ is considered as
\begin{equation}
p\left( {\bm{\alpha}  ;{a_0},{b_0}} \right) = \prod\limits_{i = 1}^N \Gamma  \left( {{\alpha _i}|{a_0},{b_0}} \right).
\label{eq15}
\end{equation}
The overall prior $p\left( \bm{\omega}  \right)$ can be obtained by computing the marginal integral of hyper-parameters in $\bm{\alpha}$ as
\begin{equation}
p\left( \bm{\omega}  \right) = \int {p\left( {\bm{\omega} |\bm{\alpha} } \right)p\left( \bm{\alpha}  \right)d} \bm{\alpha}  ,
\label{eq16}
\end{equation}
\begin{equation}
\begin{aligned}
p\left( {{\omega _i}} \right) &= \int {p\left( {{\omega _i}|{\alpha _i}} \right)p\left( {{\alpha _i}} \right)d} {\alpha _i} \\ 
&= \frac{{b_0^{{a_0}}\Gamma \left( {{a_0} + {\raise0.7ex\hbox{$1$} \!\mathord{\left/
					{\vphantom {1 2}}\right.\kern-\nulldelimiterspace}
				\!\lower0.7ex\hbox{$2$}}} \right)}}{{\sqrt {2\pi } \Gamma \left( {{a_0}} \right)}}\left( {{b_0} + {\raise0.7ex\hbox{${\omega _i^2}$} \!\mathord{\left/
			{\vphantom {{\omega _i^2} 2}}\right.\kern-\nulldelimiterspace}
		\!\lower0.7ex\hbox{$2$}}} \right)_{}^{ - \left( {{a_0} + {\raise0.7ex\hbox{$1$} \!\mathord{\left/
				{\vphantom {1 2}}\right.\kern-\nulldelimiterspace}
			\!\lower0.7ex\hbox{$2$}}} \right)}. \\ 
\end{aligned} 
\label{eq17}
\end{equation}
Since the integral is computable for Gamma $p\left( \bm{\alpha}  \right)$, the true prior $p\left( \bm{\omega}  \right)$ in (\ref{eq17}) is a Student-t distribution. 

Following the Bayesian inference, computing the estimation of $\bm{\hat \omega} $ from the SBL probability model requires the calculation of the weight posterior of $\bm{\omega}$, which is Gaussian
\begin{equation}
\begin{aligned}
p\left( {\bm{\omega} |\bm{t},\bm{\alpha} ,\beta } \right) 
&= \frac{{p\left( {\bm{t}|\bm{\omega} ,\bm{\alpha} ,\beta } \right)p\left( {\bm{\omega} |\bm{\alpha} } \right)}}{{p\left( {\bm{t}|\bm{\alpha} ,\beta } \right)}}, \\ 
&= \mathcal{N}\left( {\bm{\omega} |\bm{\mu} ,{{\mathbf{\Sigma }}_\omega }} \right), \\ 
\end{aligned} 
\label{eq18}
\end{equation}
with
\begin{equation}
\bm{\mu}  = \beta {{\mathbf{\Sigma }}_\omega }{{\mathbf{\Phi }}^{\text{T}}}\bm{t},
\label{eq19}
\end{equation}
\begin{equation}
{{\mathbf{\Sigma }}_\omega } = {\left( {\beta {{\mathbf{\Phi }}^{\text{T}}}{\mathbf{\Phi }} + \mathcal{A}} \right)^{ - 1}},
\label{eq20}
\end{equation}
where $\mathcal{A} = {\text{diag}}\left( {{\alpha _1},{\alpha _2}, \ldots ,{\alpha _N}} \right)$. By calculating hyper-parameters $\bm{\alpha}$ and $\beta$, we can estimate the $\bm{\hat \omega} $ with the mean $\bm{\mu}$ and evaluate the recovery accuracy by the variance ${{\mathbf{\Sigma }}_\omega }$. The detailed derivations of (\ref{eq19}) and (\ref{eq20}) can be found in Appendix \ref{append1x b}.

2) GP-based channel dictionary optimization with shadow fading

Generally, correlated shadow fading has a very common basic statistical feature, i.e., nearby observations tend to be more similar than distant observations. Then, it is reasonable to assume that the nearby samples have high correlated shadow fading.

To model spatial phenomena, a prevalent methodology involves employing Gaussian Processes (GP)\cite{Ranganathan11TIM, Rasmussen05MIT} within the domain of spatial statistics. GP is a random process in a continuous domain, i.e., time and space. In a GP, each point in the continuous input space is associated with a normally distributed random variable. The signal strength distribution was effectively characterized using a Gaussian Process (GP), and the estimation of an unknown location was achieved by maximizing the joint likelihood of RSS concerning the spatial coordinates. Additionally, Gaussian Process Regression (GPR) was employed to model the intricate relationship between signal strength and location within various positioning systems\cite{He17SECON}. According to\cite{He17SECON, Wang18GLOBECOM, Dashti15Globecom}, GPs exhibit the capability to quantify uncertainty in RSS data across a continuous spatial domain, aligning with the framework of Bayesian nonparametric models. Consequently, the Gaussian process serves as a valuable tool for regressing the relationship between RSS measurement values and their corresponding spatial coordinates.

Accordingly, the spatial correlated shadow fading components of $M$ samples obey $M$-dimensional Gaussian distributions and we model the uncertainty of shadow fading as a GP
\begin{equation}
\bm{\bar \xi} _{}^s \sim \mathcal{G}\mathcal{P}\left( {{\mathbf{0}},\mathcal{C}\left( {\bm{s},\bm{s}} \right)} \right),
\label{eq21}
\end{equation}
where the mean vector is ${\mathbf{0}}$ and the covariance matrix $\mathcal{C}\left( {\bm{s},\bm{s}} \right)$ is determined by covariance function or kernel function. To guarantee the positive-definite of the covariance matrix, we choose Matérn covariance function to capture the spatial correlations of shadow fading. The Matérn covariance function is a generalization of radial basis function (RBF), which matches the physical process by allowing the random field to have great flexibility and smoothness with manageable parameter number\cite{Xu21TCCN}, as 
\begin{equation}
\mathcal{C}\left( d \right) = \sigma _{}^2\frac{{2_{}^{1 - g}}}{{\Gamma \left( g \right)}}\left( {\sqrt {2g} \frac{d}{\rho }} \right)_{}^gK_g^{}\left( {\sqrt {2g} \frac{d}{\rho }} \right),
\label{eq22}
\end{equation}
where $K_g^{}\left(  \cdot  \right)$ is the Bessel function of the second kind. $g$ and $\rho $  are the order and the non-negative  spatial decay parameter of covariance, $d$ is the distance between samples, $\sigma _{}^2$ is the marginal standard deviation controls the expected variation in the output. As the shadow fading effect is continuous in space, $g$ is not large in general and it is preferable that $g = {\raise0.7ex\hbox{$3$} \!\mathord{\left/
		{\vphantom {3 2}}\right.\kern-\nulldelimiterspace}
	\!\lower0.7ex\hbox{$2$}}$ for simplicity, where a slight error of $g$ will not affect the results substantially\cite{Xu21TCCN}. The spatial correlations of shadow fading capture the scenario characteristics, which can be utilized to mine the information in the sensing data for the inference of unsampled cubes.
Since the sparse signal $\bm{\hat \omega} $ is firstly estimated by (19), the RSS denoting path loss component can be computed for any location ${\bm{s}_m}$ as
\begin{equation}
t\left( {{\bm{s}_m}} \right) = \sum\limits_{n = 1}^N {\hat \omega _n^{}{\Phi _{m,n}}} .
\label{eq23}
\end{equation}
The shadowing effect at $\left\{ {{\bm{s}_m}} \right\}_{m = 1}^M$ (we denote it as $\bm{s}$ for the sake of presentation) is defined as
\begin{equation}
\bar \xi _{}^s = 10\log 10\left( {{\raise0.7ex\hbox{$\bm{t}$} \!\mathord{\left/
			{\vphantom {t {t\left( s \right)}}}\right.\kern-\nulldelimiterspace}
		\!\lower0.7ex\hbox{${\bm{t}\left( \bm{s} \right)}$}}} \right).
\label{eq52}
\end{equation}
Note that the logarithmic form $\bar \xi _{}^s$ is a GP and the shadow fading components $\bar \xi _{}^{{v^ * }}$ of  unsampled cubes can be further inferred by Gaussian process regression (GPR) in Section \ref{construction}.

\begin{figure*}[!t]
	\centering
	\includegraphics[width=5in]{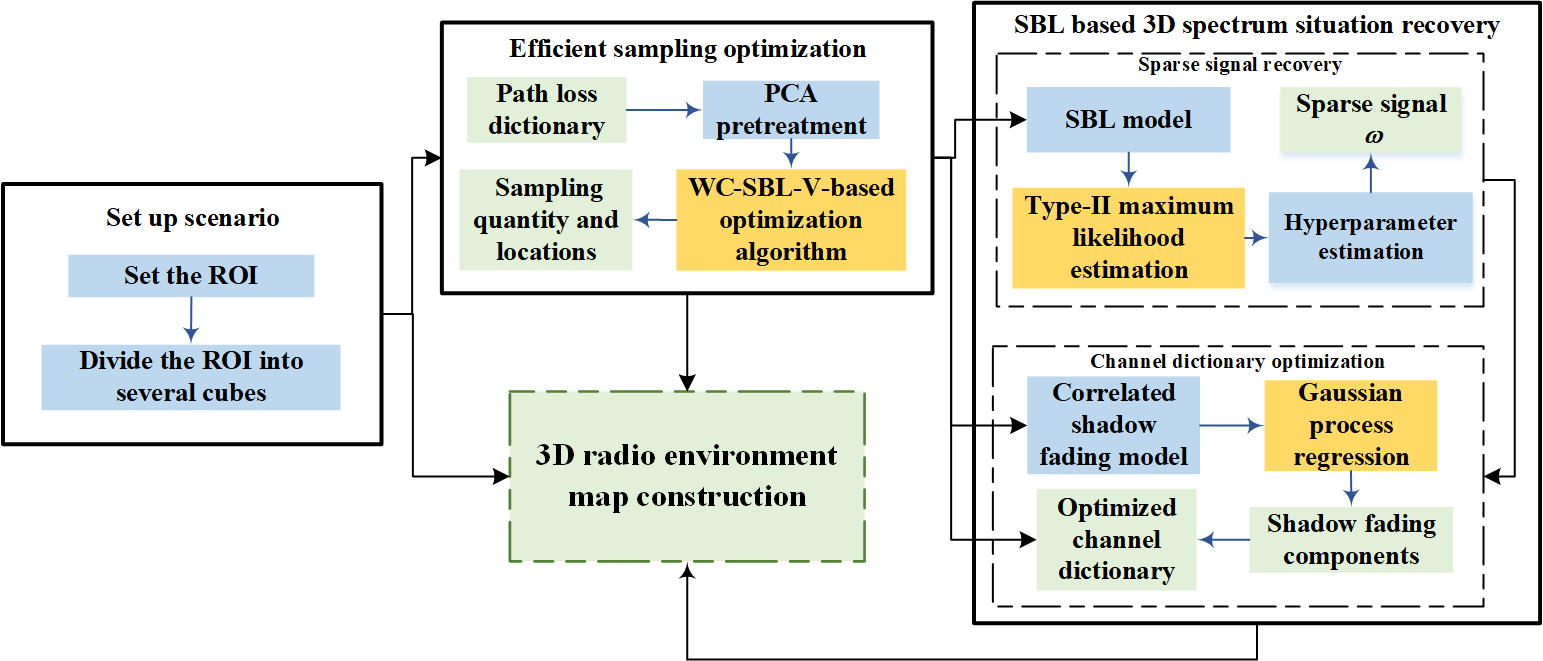}
	\caption{The flowchart of the proposed 3D REM construction scheme.}
	\label{fig2}
\end{figure*}

\section{SBL BASED 3D REM HIERARCHICAL CONSTRUCTION }
\label{SEM}
\subsection{An Overview of 3D REM Construction}
To address the spectrum data recovery problem, this section presents an SBL-based 3D REM hierarchical construction scheme. It introduces a worst case SBL variance-based measurement matrix optimization architecture to enhance the data acquisition efficiency. Moreover, the realistic propagation model and shadowing due to buildings are considered to improve the recovery accuracy. 

The flowchart of the proposed construction scheme is shown in Fig. \ref{fig2}, which mainly contains two steps, i.e., spatial measurement matrix (or sampling location) optimization and 3D spectrum data recovery. Firstly, combined with the minimum SBL variance indicator and Principal Component Analysis (PCA) dimension reduction technique, we carefully select the sampling quantity and locations. Thus, the optimized measurement matrix $\bm{\psi} $ can be obtained. Secondly, according to the propagation fading characteristics in the realistic environment, we decompose the spectrum situation recovery problem into two layers, i.e., sparse signal recovery and channel dictionary optimization. The sparse signal is firstly recovered based on the SBL by mining the inherent sparsity of spectrum situation. Then, we utilize sampling data to optimize channel dictionary by considering shadow fading. Finally, the whole REM can be constructed based on the sparse signal and optimized channel dictionary.

\subsection{Sampling Number Minimization and Location Optimization}
\label{Sampling}
According to the law of signal propagation, the 3D spectrum situation demonstrates a significant correlation in the spatial domain, i.e., the RSS values of different cubes exhibit a strong spatial correlation in the 3D space\cite{Shen22TWC}. Thus the spatial sampling scheme has a significant impact on the efficiency and accuracy of REM construction \cite{Clark19Sensors}. Compared with random or fixed sampling scheme, we can achieve better performance if the selected sampling locations can exploit the intrinsic features of spectrum data. On the other hand, it is necessary to determine the minimum number of samples and their locations to meet the required recovery accuracy. 

In the sensor network design problem, the challenge typically involves determining the optimal placement of sensor nodes. This involves identifying the minimum number of required sensor nodes and their respective sensing locations within a predefined spatial domain while ensuring that the estimation accuracy meets specified requirements\cite{Jiang16TSP}. Specifically, for a given sparse dictionary $\bm{\varphi}$, how to determine the measurement matrix $\bm{\psi} $, and thus the sparse signal can be estimated within a predefined accuracy. 

Let us define the index set $\mathcal{S}$ as the candidate sampling locations. The subset ${\mathcal{S}_d} \subset \mathcal{S}$ is defined as the determined sampling locations. According to the recovery process of sparse signal in (\ref{eq19}), it is a minimum variance unbiased estimation. The variance ${{\mathbf{\Sigma }}_\omega }$ in (\ref{eq20}) can be viewed as an error indicator to determine the sampling set $\bm{\varsigma} \left( {\bm{\varsigma}  \subset \mathcal{S}} \right)$, and the measurement matrix is determined as $\bm{\psi}  = {\left( {{{\mathbf{I}}_N}} \right)_{\bm{\varsigma} .}}$, which consists of the rows indexed by the set $\bm{\varsigma} $ in ${{\mathbf{I}}_N}$. The mean squared error (MSE) can be expressed as 
\begin{equation}
\begin{aligned}
MSE\left( {\bm{\hat \omega}} \right) &= {\text{E}}\left( {\left\| {\bm{\omega}  - \bm{\hat \omega}} \right\|_{}^2} \right) \\ 
&= {\text{tr}}\left( {{{\mathbf{\Sigma }}_\omega }} \right) \\ 
&= {\text{tr}}{\left( {\beta {{\mathbf{\Phi }}^{\text{T}}}{\mathbf{\Phi }} + \mathcal{A}} \right)^{ - 1}}, \\ 
\end{aligned} 
\label{eq24}
\end{equation}
where the sensing matrix ${\mathbf{\Phi }} = \bm{\psi}  \bm{\varphi} $ is consisted of the rows of $\varphi $ indexed by the sampling set $\varsigma $. In the initialization stage, there is no prior information of $\bm{\omega}$. We assign the same hyper-parameter $\alpha _i^{} = \alpha ,i = 1,2,...,N$. Then, we can convert (\ref{eq24}) to
\begin{equation}
\begin{aligned}
MSE\left( {\bm{\hat \omega} } \right) &= {\text{tr}}{\left( {\beta {{\mathbf{\Phi }}^{\text{T}}}{\mathbf{\Phi }} + \mathcal{A}} \right)^{ - 1}} \\ 
&= {\text{tr}}{\left[ {\beta \left( {{{\mathbf{\Phi }}^{\text{T}}}{\mathbf{\Phi }} + \frac{\alpha }{\beta }{{\mathbf{I}}_N}} \right)} \right]^{ - 1}} \\ 
&= \beta _{}^{ - 1} \cdot {\text{tr}}{\left( {{{\mathbf{\Phi }}^{\text{T}}}{\mathbf{\Phi }} + \frac{\alpha }{\beta }{{\mathbf{I}}_N}} \right)^{ - 1}}, \\ 
\end{aligned} 
\label{eq25}
\end{equation}
where ${\raise0.7ex\hbox{$\alpha $} \!\mathord{\left/
		{\vphantom {\alpha  \beta }}\right.\kern-\nulldelimiterspace}
	\!\lower0.7ex\hbox{$\beta $}}$ is the ratio of noise variance to sparse signal variance, which is sufficiently small that can be ignored \cite{Saito21Access}. Furtherly, it yields
\begin{equation}
\begin{aligned}
MSE\left( {\bm{\hat \omega} } \right) &\approx \beta _{}^{ - 1} \cdot {\text{tr}}{\left( {{{\mathbf{\Phi }}^{\text{T}}}{\mathbf{\Phi }}} \right)^{ - 1}}, \\ 
&= \beta _{}^{ - 1}\sum\limits_{n = 1}^N {{{\left( {\lambda _n^ * } \right)}^{ - 1}}} , \\ 
\end{aligned} 
\label{eq26}
\end{equation}
where $\lambda _1^ *  \geq \lambda _2^ *  \geq  \cdots  \geq \lambda _N^ * $ are the eigenvalues of $\mathbf{H} = {{\mathbf{\Phi }}^{\text{T}}}{\mathbf{\Phi }}$, here $\mathbf{H}$ is defined as the dual sensing matrix. The parameter $\beta $ is a constant and it is dependent on $\sigma _0^2$ but independent of sensing locations. 

As defined in\cite{Joshi09TSP}, the worst case error variance (WCEV), which is defined as the variance of estimation error over all directions $\bm{\ell}  \in {\mathbb{R}^N}$ with $\left\| \bm{\ell}  \right\| = 1$, can be computed as follows
\begin{equation}
\begin{aligned}
WCE{V_\mathbf{H}} &= \mathop {\max }\limits_{\left\| \bm{\ell}  \right\| = 1} \bm{\ell} _{}^{\text{T}}\mathbf{H}_{}^{ - 1}\bm{\ell}  \\ 
&= {\lambda _{\max }}\left( {\mathbf{H}_{}^{ - 1}} \right), \\ 
\end{aligned} 
\label{eq27}
\end{equation}
where $\mathbf{H}_{}^{ - 1}$ corresponds to the Fisher information matrix (FIM) and  denotes the maximum eigenvalue of $\mathbf{H}_{}^{ - 1}$, i.e., the minimum eigenvalue of $\mathbf{H}$. We denote ${\lambda _{\max }}\left( {\mathbf{H}_{}^{ - 1}} \right)$ as the WC-SBL-V index. This measure leads to the so-called E-optimal experiment design (maximum eigenvalue constraint on the FIM $\mathbf{H}_{}^{ - 1}$)\cite{Ma23TAES}. It can be proved that two error indicators are equivalent since two matrix norms, i.e., 2-norm and F-norm, are equivalent\cite{2008Matrix}. 

In order to determine the minimum sampling number, the sparse-promoting technique is used by adding a sparsity-promoting penalty term to the cost function. The maximum acceptable eigenvalue $\lambda _{WCEV}^{}$ is defined as the WC-SBL-V index threshold. Accordingly, the desired sample set $\bm{\varsigma}  \subset \mathcal{S}$ can be expressed as a sparse optimization problem, 
\begin{equation}
\begin{gathered}
\bm{\hat \varsigma}  = \arg \mathop {\min }\limits_{\bm{\varsigma}  \subset \mathcal{S}} \left| \bm{\varsigma}  \right|, \\ 
{\text{s}}{\text{.t}}{\text{. }}{\lambda _{\max }}\left( {\mathbf{H}_{}^{ - 1}} \right) \geq \lambda _{WCEV}^{}, \\ 
\mathbf{H} = {{\mathbf{\Phi }}^{\text{T}}}{\mathbf{\Phi }}, \\ 
{\mathbf{\Phi }} = \bm{\psi} \bm{\varphi} , \\ 
\bm{\psi}  = {\left( {{{\mathbf{I}}_N}} \right)_{\bm{\varsigma}  \cdot }}, \\ 
\end{gathered} 
\label{eq28}
\end{equation}
where $\left|  \cdot  \right|$ is the cardinality of a set. 

The cardinality optimization problem of (\ref{eq28}) can be solved by evaluating the maximum eigenvalue of Fisher information matrix over all potential sampling location configuration. One straightforward method is to evaluate the performance of all possible combinations of the potential sizes of the candidate sensing locations. The location with the least number of sensor nodes that satisfies the required estimation accuracy is chosen.. The computational cost of exhaustively searching is ${2^N}$, which is computationally intractable for a large-scale problem.

One effective approach to reduce the search space for sensor configurations is to determine the sensing locations incrementally. Accordingly, the greedy algorithm is adopted in this paper. Each sampling location is selected after traversing all the unselected samples and find the one that maximizes the minimum nonzero eigenvalue. However, the candidate observation matrix has full row rank (${\mathbf{\Phi }} \in {\mathbb{R}^{M \times N}},M < N$). Therefore, the dimension of $\mathbf{H}$ is ${\mathbb{R}^{N \times N}}$, but it is computationally expensive to calculate the eigenvalues due to the high spatial dimension. PCA is an efficient dimension reduction technology, and it is utilized to reduce the dictionary from $N$ vectors to $n$ vectors, which are the principle components of the  $N$-colum vectors of $\bm{\varphi}  \in {\mathbb{R}^{N \times N}}$, and we can obtain $\bm{\varphi} _{}^p \in {\mathbb{R}^{N \times n}}$. The corresponding dual observation matrix can be converted to

\begin{equation}
\mathbf{H}_{}^p = {\left( {{\mathbf{\Phi }}_{}^p} \right)^{\text{T}}}{\mathbf{\Phi }}_{}^p,
\label{eq29}
\end{equation}
where ${\mathbf{\Phi }}_{}^p$ is consisted of the rows indexed by set $\varsigma $ in $\bm{\varphi} _{}^p$, and $\lambda _1^{} \geq \lambda _2^{} \geq  \cdots  \geq \lambda _n^{}$ are the eigenvalues of $\mathbf{H}_{}^p$. Then, let ${\mathbf{\Phi }}_t^p$ denote the sensing matrix after selecting $t$ samples as
\begin{equation}
{\mathbf{\Phi }}_t^p = \left[ {{{\left( {\varphi _{{i_1}}^p} \right)}^{\text{T}}},{{\left( {\varphi _{{i_2}}^p} \right)}^{\text{T}}}, \cdots ,{{\left( {\varphi _{{i_{t - 1}}}^p} \right)}^{\text{T}}},{{\left( {\varphi _{{i_t}}^p} \right)}^{\text{T}}}} \right]_{}^{\text{T}},
\label{eq30}
\end{equation}
where ${i_t}\left( {{i_t} \in \varsigma ,{i_t} \in \mathcal{S}} \right)$ is the index of $t\text{th}$ selected sample. The observation vector $\varphi _{{i_t}}^p$ is the ${i_t}\text{th}$ row vector of $\bm{\varphi} _{}^p$. Accordingly, the sparse optimization problem of (\ref{eq28}) can be converted to
\begin{equation}
\begin{gathered}
\bm{\hat \varsigma}  = \arg \mathop {\min }\limits_{\bm{\varsigma}  \subset \mathcal{S}} \left| \bm{\varsigma}  \right|, \\ 
{\text{s}}{\text{.t}}{\text{. }}\lambda _n^{} \geq \lambda _{WCEV}^{}, \\ 
\mathbf{H}_{}^p = {\left( {{\mathbf{\Phi }}_{}^p} \right)^{\text{T}}}{\mathbf{\Phi }}_{}^p, \\ 
{\mathbf{\Phi }}_{}^p = \bm{\varphi} _{\bm{\varsigma}  \cdot }^p. \\ 
\end{gathered} 
\label{eq31}
\end{equation}

Let $\lambda _n^{}\left( {
	\mathbf{H}_{t - 1}^p} \right)$ denote the minimum eigenvalue of $\mathbf{H}_{t - 1}^p$ as
\begin{equation}
\lambda _n^{}\left( {\mathbf{H}_{t - 1}^p} \right) = \mathop {\min }\limits_{\left\| \bm{\ell}  \right\| = 1} \bm{\ell} _{}^{\text{T}}\mathbf{H}_{t - 1}^p\bm{\ell} .
\label{eq32}
\end{equation}
After the $t\text{th}$ sample is selected, the minimum eigenvalue of $\mathbf{H}_t^p$ can be expressed as
\begin{equation}
\lambda _n^{}\left( {\mathbf{H}_t^p} \right) = \mathop {\min }\limits_{\left\| \bm{\ell}  \right\| = 1,{i_t} \in \mathcal{S}\backslash {\mathcal{S}_d}} \bm{\ell} _{}^{\text{T}}\mathbf{H}_{t - 1}^p\bm{\ell}  + {\left( {{{\left( {\varphi _{{i_t}}^p} \right)}^{\text{T}}}\bm{\ell} } \right)^2},
\label{eq33}
\end{equation}
with
\begin{equation}
\bm{\ell}  = \bm{\pi} _n^{\left( {t - 1} \right)}\mathbf{H}_{t - 1}^p,
\label{eq34}
\end{equation}
where $\bm{\pi} _n^{\left( {t - 1} \right)}$ is the normalized minimum eigenvector of $\mathbf{H}_{t - 1}^p$. Moreover, the minimum eigenspace is also introduced here via optimizing the new criterion efficiently\cite{Jiang16TSP}. The $t\text{th}$ sample can be determined by maximizing the projection of row vector $\varphi _{{i_t}}^p$ on the minimum eigenspace of $\mathbf{H}_{t - 1}^p$. 

In the process of determining $\bm{\varsigma} $, the number of samples is different due to the different threshold value. Then, $t$ can appear in two cases, i.e., $t \leq n$ and $t > n$. The minimum eigenspace of $\mathbf{H}_{t - 1}^p$ is defined for the two cases as follows.

i) For $t \leq n$, the minimum eigenspace of $\mathbf{H}_{t - 1}^p$ is the eigenspace associated with all the minimum eigenvalues of $\mathbf{H}_{t - 1}^p$, as
\begin{equation}
{\mathbf{\Pi }}_{t:n}^{t - 1}\left( {\mathbf{H}_{t - 1}^p} \right) = {\text{span}}\left( {{\bm{\pi}}_t^{\left( {t - 1} \right)},{\bm{\pi}}_{t + 1}^{\left( {t - 1} \right)}, \ldots ,{\bm{\pi}}_n^{\left( {t - 1} \right)}} \right),
\label{eq35}
\end{equation}
where ${\bm{\pi}}_j^{\left( {t - 1} \right)}$ is the normalized eigenvector associated with the $j\text{th}$ eigenvalue ${\lambda _j}\left( {\mathbf{H}_{t - 1}^p} \right)$ of $\mathbf{H}_{t - 1}^p$. The minimum eigenspace of $\mathbf{H}_{t - 1}^p$ of (\ref{eq35}) is equal to the null space of $\mathbf{H}_{t - 1}^p$ as
\begin{equation}
{\mathbf{\Pi }}_{t:n}^{t - 1}\left( {\mathbf{H}_{t - 1}^p} \right){\text{ =  null}}\left( {{\mathbf{\Phi }}_{t - 1}^p} \right).
\label{eq36}
\end{equation}

ii) For $t > n$, the minimum eigenspace of $\mathbf{H}_{t - 1}^p$ is exactly the subspace spanned by the minimum eigenvector $\pi _n^{\left( {t - 1} \right)}$ as
\begin{equation}
{\mathbf{\Pi }}_n^{t - 1}\left( {\mathbf{H}_{t - 1}^p} \right) = {\text{span}}\left( {{\bm{\pi}}_n^{\left( {t - 1} \right)}} \right) = \Pi _n^{t - 1}.
\label{eq37}
\end{equation}	
When $t \leq n$, the $t\text{th}$ sample can be determined by
\begin{equation}
{\hat i_t} = \arg \mathop {\max }\limits_{{i_t} \in \mathcal{S}\backslash {\mathcal{S}_d}} \left\| {{\mathcal{O}_{t - 1}}\bm{\varphi} _{{i_t}}^p} \right\|_2^2,
\label{eq38}
\end{equation}	
with
\begin{equation}
{\mathcal{O}_{t - 1}} = {{\mathbf{I}}_n} - {\Theta _{t - 1}}\Theta _{t - 1}^T,
\label{eq39}
\end{equation}
where ${\Theta _{t - 1}} = {\text{orth}}\left( {\left( {{\mathbf{\Phi }}_{t - 1}^p} \right)_{}^{\text{T}}} \right)$. ${\mathcal{O}_{t - 1}}$ is a projection matrix which can project $\bm{\varphi} _{{i_t}}^p$ onto the minimum eigenspace ${\text{null}}\left( {{\mathbf{\Phi }}_{t - 1}^p} \right)$. When $t > n$, the $t\text{th}$ sample can be obtained by
\begin{equation}
{\hat i_t} = \arg \mathop {\max }\limits_{{i_t} \in \mathcal{S}\backslash {\mathcal{S}_k}} \left\| {{\mathcal{R}_{t - 1}}\bm{\varphi} _{{i_t}}^p} \right\|_2^2,
\label{eq40}
\end{equation}	
with
\begin{equation}
{\mathcal{R}_{t - 1}} = {\bm{\pi}}_n^{t - 1}\left( {{\bm{\pi}}_n^{t - 1}} \right)_{}^{\text{T}},
\label{eq41}
\end{equation}	
where ${\mathcal{R}_{t - 1}}$ is a projection matrix which can project $\bm{\varphi} _{{i_t}}^p$ onto the minimum eigenspace $\mathbf{\Pi} _n^{t - 1}$. 

We examine all of the unchosen observation vectors and select the one that maximizes the eigenspace. Meanwhile, if $t \geq n$, we check the constraint in (\ref{eq31}) after each sampling location is determined. If the constraint is satisfied, stop the algorithm. That is, the process of (\ref{eq36})-(\ref{eq41}) is iterated until the cut-off condition $\lambda _n^{} \geq \lambda _{WCEV}^{}$ is satisfied. Then, the minimum and optimized sampling set $\bm{\varsigma}  = \left[ {{i_1},{i_2}, \ldots ,{i_M}} \right]$ can be obtained. It should be mentioned that the minimum number of required samples is determined by judging whether $\lambda _n^{(t)} \geq \lambda _{WCEV}^{}$ is satisfied after the $t\text{th}$ sample is determined. Accordingly, the constraint is only used to judge whether the number of required samples is enough. Algorithm \ref{alg1} summarizes the process of optimization. 

The time complexity of sparse dictionary PCA preprocessing is $O\left( {N_{}^3} \right)$. The rest time complexity mainly focuses on line 3 and line 15 in Algorithm \ref{alg1}. To determine the $t\text{th}$ sample, the main computational cost is attributed to the optimization problem ${\hat i_t} = \arg \mathop {\max }\limits_{{i_t} \in \mathcal{S}\backslash {\mathcal{S}_d}} \left\| {{\mathcal{O}_{t - 1}}\bm{\varphi} _{{i_t}}^p} \right\|_2^2$ and ${\hat i_t} = \arg \mathop {\max }\limits_{{i_t} \in \mathcal{S}\backslash {\mathcal{S}_k}} \left\| {{\mathcal{R}_{t - 1}}\bm{\varphi} _{{i_t}}^p} \right\|_2^2$, which cost $O\left( {\left( {N - t + 1} \right)n_{}^2} \right)$ per selection. The selection of required $M$ samples under the given threshold total costs $O\left( {NMn_{}^2} \right)$. Then, the total time complexity ${\mathcal{C}_1}$ of Algorithm \ref{alg1} is
\begin{equation}
\begin{gathered}
{\mathcal{C}_1} = O\left( {N_{}^3} \right) + O\left( {\sum\limits_{t = 1}^M {\left( {N - t + 1} \right)n_{}^2} } \right) \\ 
= O\left( {N_{}^3 + NMn_{}^2} \right). \\ 
\end{gathered} 
\label{eq42}
\end{equation}	

\begin{algorithm}[htb] 
	\caption{ WC-SBL-V based measurement matrix optimization.} 
	\label{alg1} 
	\begin{algorithmic}[1] 
		\REQUIRE ~~\\ 
		3D REM sparse dictionary $\bm{\varphi}  \in {\mathbb{R}^{N \times N}}$; \\
		Initial selected sample set $\bm{\varsigma}  = \not 0$; \\
		The maximum acceptable variance $\lambda _{WCEV}^{}$;\\
		\ENSURE ~~\\ 
		Selected sample set $\bm{\varsigma} $; The measurement matrix $\bm{\psi} $; The number of selected samples $M$;
		\STATE \textbf{Initialize} $\mathcal{S} = \left\{ {1,2, \ldots ,N} \right\}$, ${\mathcal{S}_d} = \not 0$, $t = 0$, ${\lambda _n} = \infty $, ${\mathcal{O}_0} = {\mathbf{I}_n}$; 
		\label{ initialize }
		\STATE Obtain $\varphi _{}^p \in {\mathbb{R}^{N \times n}}$ by PCA; 
		\label{code:fram:trainbase}
		\STATE \textbf{while} $t \leq n$ \& $\lambda _n^{(t)} \geq \lambda _{WCEV}^{}$ do; 
		\label{code:fram:add}
		\STATE $t = t + 1$; 
		\label{code:fram:classify}
		\STATE Solve ${\hat i_t}$ according to (\ref{eq38}); 
		\label{code:fram:select}
		\STATE Update $\bm{\varsigma}  = \bm{\varsigma}  \cup \left\{ {{{\hat i}_t}} \right\}$, ${\mathbf{\Phi }}_t^p = {\left[ {{{\left( {{\mathbf{\Phi }}_{t - 1}^p} \right)}^{\text{T}}}\varphi _{{{\hat i}_t}}^{\text{p}}} \right]^{\text{T}}}$, ${\Theta _t} = {\text{orth}}\left( {\left( {{\mathbf{\Phi }}_t^p} \right)_{}^{\text{T}}} \right)$, ${\mathcal{O}_t} = {{\mathbf{I}}_n} - {\Theta _t}\Theta _t^T$; 
		\STATE Establish the projection matrix ${\mathcal{O}_t}$ on the minimum eigenspace according to (\ref{eq39}); 
		\STATE ${\mathcal{S}_d} = \bm{\varsigma} $; 
		\STATE \textbf{end while}; 
		\STATE \textbf{while} $t > n$ \& $\lambda _n^{(t)} \geq \lambda _{WCEV}^{}$ do;
		\STATE $t = t + 1$;  
		\STATE Solve ${\hat i_t}$ according to (\ref{eq40}); 
		\STATE Update $\bm{\varsigma}  = \bm{\varsigma}  \cup \left\{ {{{\hat i}_t}} \right\}$, ${\mathbf{\Phi }}_t^p = {\left[ {{{\left( {{\mathbf{\Phi }}_{t - 1}^p} \right)}^{\text{T}}}\varphi _{{{\hat i}_t}}^{\text{p}}} \right]^{\text{T}}}$, ${\left( {{\mathbf{\Phi }}_t^p} \right)^{\text{T}}}{\mathbf{\Phi }}_t^p = {\mathbf{\Pi }}_{}^t\lambda _{}^{\left( t \right)}{\left( {{\mathbf{\Pi }}_{}^t} \right)^{\text{T}}}$, $\bm{\lambda} _{}^{\left( t \right)} = diag\left[ {\lambda _1^{(t)},\lambda _2^{(t)}, \ldots ,\lambda _n^{(t)}} \right]$, ${\mathcal{R}_t} = {\bm{\pi }}_n^t\left( {{\bm{\pi }}_n^t} \right)_{}^{\text{T}}$; 
		\STATE Obtain the minimum eigenspace according to (\ref{eq37}); 
		\STATE Establish the projection matrix ${\mathcal{R}_t}$ on the minimum eigenspace according to (\ref{eq41}); 
		\STATE ${\mathcal{S}_d} = \bm{\varsigma} $; 
		\STATE \textbf{end while}; 
		\RETURN $\bm{\varsigma} $, $M = t$, $\bm{\psi}  = {\left( {{{\mathbf{I}}_N}} \right)_{\varsigma }}$; 
	\end{algorithmic}
\end{algorithm}

\subsection{SBL-Based Spectrum Data Recovery Incorporating Channel Shadowing}
\label{construction}
As discussed in Section \ref{SSM}, the REM data is recovered hierarchically by SBL and GPR. Given the sampling data, we estimate the sparse RF transmitter signal $\bm{\hat \omega} $ by SBL. We further derive the shadow fading components at the sampling locations and construct a GP to estimate the channel shadowing $\bm{\bar \xi} _{}^{{v^ * }}$ at the unsampled locations. Finally, we recover the spectrum data according to (\ref{eq2}).

Firstly, we can recover $\bm{\omega} $ with the mean $\mu $ and evaluate the recovery accuracy by the variance ${{\mathbf{\Sigma }}_\omega }$ once the hyper-parameters $\bm{\alpha} $ and $\beta $ have been estimated, which are estimated by a maximum a posterior (MAP) probability as\cite{Tipping01JMLR}

\begin{equation}
\begin{aligned}
\left( {\bm{\alpha} ,\beta } \right) &= \mathop {\arg \max }\limits_{\bm{\alpha} ,\beta } p\left( {\bm{\alpha} ,\beta |\bm{t}} \right), \\ 
&= \mathop {\arg \max }\limits_{\bm{\alpha} ,\beta } p\left( {\bm{t}|\bm{\alpha} ,\beta } \right)p\left( \bm{\alpha}  \right)p\left( \beta  \right), \\ 
&= \mathop {\arg \max }\limits_{\bm{\alpha} ,\beta } \ln p\left( {\bm{t}|\bm{\alpha} ,\beta } \right)p\left( \bm{\alpha}  \right)p\left( \beta  \right). \\ 
\end{aligned} 
\label{eq43}
\end{equation}	
The observations determine these parameters by approximating the hyperparameter posterior in (\ref{eq18}) by its mode. This is equal to maximizing the evidence for the measurements $p\left( {\bm{t}|\bm{\alpha} ,\beta } \right)$ in (\ref{eq43}), which is often referred to as type-II maximum likelihood, intuitively selecting the hyperparameters that are best supported by the observations\cite{Kiaee16TNNLS}.

Then, the re-estimation rule of $\alpha _i^{}$ is \cite{Zhang14TNSRE}
\begin{equation}
\alpha _i^ *  = \frac{{1 + 2a}}{{\mu _i^2 + {{\left( {{\Sigma _\omega }} \right)}_{i,i}} + 2b}}.
\label{eq44}
\end{equation}
However, by defining quantities ${\Upsilon _i} \equiv 1 - {\alpha _i}{\Sigma _{ii}}$, it yields
\begin{equation}
\alpha _i^ *  = \frac{{{\Upsilon _i} + 2a}}{{\mu _i^2 + 2b}},
\label{eq45}
\end{equation}
which can lead to much faster convergence than (\ref{eq44}). The update rule of $\beta $ can be expressed as 
\begin{equation}
\beta _{}^ *  = \frac{{M - \sum\limits_i {{\Upsilon _i}}  + 2c}}{{\left\| {\bm{t} - {\mathbf{\bm{\Phi} }}\mu } \right\|_2^2 + 2d}}.
\label{eq46}
\end{equation}	
The derivations of (\ref{eq44}) and (\ref{eq46}) are given in Appendix \ref{append1x c}.

If any $\alpha _i^{ - 1} = 0\left( {{\alpha _i} \to \infty } \right)$, the corresponding ${\omega _i} = 0$ and the targets are unlikely to locate in the $i\text{th}$ cube, i.e., the transmitter cannot be in the $i\text{th}$ cube\cite{Tipping01JMLR}. Therefore, we can remove these locations to accelerate the update process. Since the matrix inversion of SBL variance ${{\mathbf{\Sigma }}_\omega }$ in (\ref{eq20}) is also computationally complex, we apply the matrix inversion lemma and write it as
\begin{equation}
{{\mathbf{\Sigma }}_\omega } = \left( \mathcal{A} \right)_{}^{ - 1} - \left( \mathcal{A} \right)_{}^{ - 1}{{\mathbf{\Phi }}^{\text{T}}}\left( \Omega  \right)_{}^{ - 1}{\mathbf{\Phi }}\left( \mathcal{A} \right)_{}^{ - 1},
\label{eq50}
\end{equation}
with
\begin{equation}
\Omega  = {\mathbf{\Phi }}\left( \mathcal{A} \right)_{}^{ - 1}{{\mathbf{\Phi }}^T} + \beta _{}^{ - 1}{\mathbf{I}}.
\label{eq51}
\end{equation}
Note that its computational complexity is lowered than $O\left( {{M^2}N} \right)$ while the one of (\ref{eq20}) is $O\left( {{N^3}} \right)$. By performing the iteration between (\ref{eq45}), (\ref{eq46}), (\ref{eq50}), (\ref{eq51}) and (\ref{eq19}) until the convergence condition is satisfied, the MAP estimation of $\bm{\omega} $ can be obtained from the mean of posterior, i.e., $\bm{\hat \omega}  = \mu $.

Secondly, according to correlated shadow fading model in Section \ref{SSM}, we introduce the shadow fading estimation based on the GPR. In general, noise is considered in the GP model. With training the input set $\bm{\bar \xi} _{}^s$ calculated by (\ref{eq52}), the corresponding model output is defined as
\begin{equation}
{\mathbf{y}} = \bm{\bar \xi} _{}^s + \bm{\delta} ,
\label{eq53}
\end{equation}
where $\begin{array}{*{20}{c}}
{{\delta _m} \sim \mathcal{N}\left( {0,\sigma _{GP}^2} \right),}&{m = 1,2, \ldots ,M} 
\end{array}.$ The joint distribution of the prediction values $\bar \xi _{}^{{v^ * }}$ at the unsampled cubes ${\mathbf{\bm{\nu} }}_{}^ * \left( {{\mathbf{\bm{\nu} }}_{}^ *  = \left\{ {i|i \in {\mathbf{\bm{\nu} }},i \notin \bm{s}} \right\}} \right)$ can be represented as a multi-normal distribution
\begin{equation}
\left[ {\begin{array}{*{20}{c}}
	{\mathbf{y}} \\ 
	{\bm{\bar \xi} _{}^{{v^ * }}} 
	\end{array}} \right]|s,{\mathbf{\bm{\nu} }}_{}^ *  \sim \mathcal{N}\left( {\left[ {\begin{array}{*{20}{c}}
		{\mathbf{0}} \\ 
		{\mathbf{0}} 
		\end{array}} \right],\left[ {\begin{array}{*{20}{c}}
		{\mathcal{C}\left( {\bm{s},\bm{s}} \right) + \sigma _{GP}^2{\mathbf{I}}}&{\mathcal{C}\left( {\bm{s},{\mathbf{\bm{\nu} }}_{}^ * } \right)} \\ 
		{\mathcal{C}\left( {{\mathbf{\bm{\nu} }}_{}^ * ,\bm{s}} \right)}&{\mathcal{C}\left( {{\mathbf{\bm{\nu} }}_{}^ * ,{\mathbf{\bm{\nu} }}_{}^ * } \right)} 
		\end{array}} \right]} \right).
\label{eq54}
\end{equation}
According to GPR, the predictive distribution $\bm{\bar \xi} _{}^{{v^ * }}$ satisfies the multivariate Gaussian distribution
\begin{equation}
p\left( {\bm{\bar \xi} _{}^{{v^ * }}|{\mathbf{y}},\bm{s},{\bm{\mathbf{\nu }}}_{}^ * } \right) \sim \mathcal{N}\left( {\bm{\mu} _{GP}^ * ,{\mathbf{\Sigma }}_{GP}^ * } \right),
\label{eq55}
\end{equation}
with
\begin{equation}
\bm{\mu} _{GP}^ *  = \mathcal{C}\left( {{\bm{\mathbf{\nu }}}_{}^ * ,\bm{s}} \right)\left( {\mathcal{C}\left( {\bm{s},\bm{s}} \right) + \sigma _{GP}^2{\mathbf{I}}} \right)_{}^{ - 1}{\mathbf{y}},
\label{eq56}
\end{equation}
\begin{equation}
\begin{aligned}
{\mathbf{\Sigma }}_{GP}^ *  &= \mathcal{C}\left( {{\bm{\mathbf{\nu }}}_{}^ * ,{\bm{\mathbf{\nu }}}_{}^ * } \right) + \sigma _{GP}^2{\mathbf{I}}\\
& - \mathcal{C}\left( {{\bm{\mathbf{\nu }}}_{}^ * ,\bm{s}} \right)\left({\mathcal{C}\left( {\bm{s},\bm{s}} \right) + \sigma _{GP}^2{\mathbf{I}}} \right)_{}^{ - 1}\mathcal{C}\left( {\bm{s},{\bm{\mathbf{\nu }}}_{}^ * } \right). \\
\end{aligned}
\label{eq57}
\end{equation}

We apply Bayesian inference idea to solve GPR. The marginal likelihood is given by
\begin{equation}
p\left( {{\mathbf{y}}|\bm{s},{\eta _{GP}}} \right) \sim \mathcal{N}\left( {{\mathbf{y}}|{\mathbf{0}},{{\mathbf{\Sigma }}_{{\eta _{GP}}}}} \right),
\label{eq58}
\end{equation}
\begin{equation}
{{\mathbf{\Sigma }}_{{\eta _{GP}}}} = \mathcal{C}\left( {\bm{s},\bm{s}} \right) + \sigma _{GP}^2{\mathbf{I}},
\label{eq59}
\end{equation}
where ${\eta _{GP}} = [\rho ,\sigma _{}^2,\sigma _{GP}^2]$ are the parameters of the Matérn covariance function and GP noise variance. The parameters ${\eta _{GP}}$ can be estimated by minimizing the negative log marginal likelihood (NLML) with respect to ${\eta _{GP}}$, as 
\begin{equation}
\begin{aligned}
{{\hat \eta }_{GP}} &= \mathop {\arg \min }\limits_{{\eta _{GP}}} \mathcal{L}\left( {{\eta _{GP}}} \right), \\ 
&= \mathop {\arg \min }\limits_{{\eta _{GP}}} \left( { - \log p\left( {{\mathbf{y}}|\bm{s},{\eta _{GP}}} \right)} \right), \\ 
&= \mathop {\arg \min }\limits_{{\eta _{GP}}} \left( {\frac{1}{2}{\mathbf{y}}_{}^{\text{T}}{\mathbf{\Sigma }}_{{\eta _{GP}}}^{ - 1}{\mathbf{y}} + \frac{1}{2}\log \left( {\left| {{\mathbf{\Sigma }}_{{\eta _{GP}}}^{}} \right|} \right) + \frac{M}{2}\log 2\pi } \right). \\ 
\end{aligned} 
\label{eq60}
\end{equation}
which is a non-convex problem. The optimum of problem (\ref{eq60}) can be solved by the gradient-based optimization algorithm
\begin{equation}
\begin{aligned}
{\raise0.7ex\hbox{${\partial \mathcal{L}\left( {{\eta _{GP}}} \right)}$} \!\mathord{\left/
		{\vphantom {{\partial \mathcal{L}\left( {{\eta _{GP}}} \right)} {\partial {\eta _{GP}}}}}\right.\kern-\nulldelimiterspace}
	\!\lower0.7ex\hbox{${\partial {\eta _{GP}}}$}} &=  - \frac{1}{2}{\mathbf{y}}_{}^{\text{T}}{\mathbf{\Sigma }}_{{\eta _{GP}}}^{ - 1}\frac{{\partial {\mathbf{\Sigma }}_{{\eta _{GP}}}^{}}}{{\partial {\eta _{GP}}}}{\mathbf{\Sigma }}_{{\eta _{GP}}}^{ - 1}{\mathbf{y}}\\
& + \frac{1}{2}{\text{tr}}\left( {{\mathbf{\Sigma }}_{{\eta _{GP}}}^{ - 1}\frac{{\partial {\mathbf{\Sigma }}_{{\eta _{GP}}}^{}}}{{\partial {\eta _{GP}}}}} \right). \\
\end{aligned} 
\label{eq61}
\end{equation}

After obtaining the prediction model $p\left( {\bm{\bar \xi} _{}^{{v^ * }}|{\mathbf{y}},\bm{s},{\bm{\mathbf{\nu }}}_{}^ * } \right)$, the shadow fading components $\bm{\bar \xi} _{}^{{v^ * }}$ of unsampled cubes can be estimated by $\bm{\mu} _{GP}^ * $. We further obtain shadow fading $\bm{\xi} _{}^v = {10^{{\raise0.7ex\hbox{${\bm{\bar \xi} _{}^v}$} \!\mathord{\left/
				{\vphantom {{\bm{\bar \xi} _{}^v} {10}}}\right.\kern-\nulldelimiterspace}
			\!\lower0.7ex\hbox{${10}$}}}}$ of all cubes. Finally, we can reconstruct the 3D REM (or REM tensor) by using (\ref{eq2}). The recovery process is summarized in Algorithm \ref{alg2}. 
		
		The complexity of SBL-based REM construction is mainly concentrated in line 3, 4 and 7 in Algorithm \ref{alg2}. Since the iteration number is a constant, this algorithm’s complexity is decided by the most time-consuming step. As discussed above, the computation of (\ref{eq50}) costs $O\left( {{M^2}N} \right)$. The computation of (\ref{eq19}) and (\ref{eq46}) both cost $O\left( {MN_{}^2} \right)$, which occupy the largest proportion of computational cost. Therefore, the total time complexity of SBL for sparse signal recovery is $O\left( {MN_{}^2} \right)$. The time complexity of GPR for shadow fading estimation mainly lies in the process of Bayesian inference with $M$ samples, which is $O\left( {M_{}^3} \right)$. Accordingly, the total complexity is
\begin{equation}
\begin{aligned}
{\mathcal{C}_2} &= O\left( {MN_{}^2} \right) + O\left( {M_{}^3} \right) \\ 
&= O\left( {M_{}^3 + MN_{}^2} \right). \\ 
\end{aligned}
\label{eq62}
\end{equation}

\begin{algorithm}[htbp] 
	\caption{ SBL-based spectrum data hierarchical recovery incorporating channel shadowing.} 
	\label{alg2} 
	\begin{algorithmic}[1] 
		\REQUIRE ~~\\ 
		3D REM sensing matrix ${\mathbf{\Phi }} \in \mathbb{R}_{}^{M \times N}$;\\
		3D REM sparse dictionary $\bm{\varphi}  \in {\mathbb{R}^{N \times N}}$; \\
		RSS $\bm{t}$; 
		The locations of samples $\bm{s}$;  
		$thre\_\alpha $; $a$; $b$; $c$; $d$; \\
		Max iteration $ite{r_{\max }}$; \\
		The optimized spatial measurement matrix $\bm{\psi} $;
		\ENSURE ~~\\ 
		Constructed 3D REM vector ${\mathbf{x}}$; Recovered sparse signal $\bm{\hat \omega} $; The estimated shadow fading $\bm{\xi} _{}^v$;
		\STATE \textbf{Initialize} $\bm{\alpha} $, $\beta $, $iter = 1$, $\bm{\mu}  = {{\mathbf{0}}_{N \times 1}}$, ${{\mathbf{\Sigma }}_\omega }$; 
		\STATE \textbf{while} $iter < ite{r_{\max }}$ or \\
	    $\left\| {\mathcal{L}\left( {{\alpha ^{iter}},{\beta ^{iter}}} \right) - \mathcal{L}\left( {{\alpha ^{iter - 1}},{\beta ^{iter - 1}}} \right)} \right\| < {10^{ - 4}}$ do; 
		\STATE Update $\bm{\alpha} $ with (\ref{eq45}); 
		\STATE Update $\beta $ with (\ref{eq46}); 
		\STATE Calculate $\bm{\mu}$ and ${{\mathbf{\Sigma }}_\omega }$ by (\ref{eq19}), (\ref{eq50}) and (\ref{eq51}); 
		\STATE $iter = iter + 1$;
		\STATE \textbf{end while}; 
		\STATE Obtain $\bm{\hat \omega}  = \bm{\mu} $; 
		\STATE Calculate the shadow fading at $\bm{s}$ by (\ref{eq52}); 
		\STATE Calculate the NLML $\mathcal{L}\left( {{\eta _{GP}}} \right)$ by (\ref{eq58}) - (\ref{eq60}); 
		\STATE Solve ${\hat \eta _{GP}} = \mathop {\arg \min }\limits_{{\eta _{GP}}} \mathcal{L}\left( {{\eta _{GP}}} \right)$ based on (\ref{eq61}); 
		\STATE Get prediction model $p\left( {\bm{\bar \xi} _{}^{{v^ * }}|{\mathbf{y}},\bm{s},{\bm{\nu }}_{}^ * } \right)$ by (\ref{eq55}) - (\ref{eq57}); 
		\STATE Obtain $\bm{\bar \xi} _{}^{{v^ * }} = \mu _{GP}^ * $, ${\mathbf{x}} = \bm{\xi} _{}^v \circ \left( {\bm{\varphi} \bm{\hat \omega} } \right)$; 
	\end{algorithmic}
\end{algorithm}

\section{Simulation Results And Discussions}
\subsection{Experiment Setup}
In this section, the proposed 3D REM construction method is validated and evaluated by simulations. The 3D ROI is a campus scenario and the size is $1250{\text{m}} \times 1250{\text{m}} \times 50{\text{m}}{\text{.}}$ The satellite view is shown in Fig. \ref{fig3} (a). There are many buildings densely distributed with heights from 19 to 55m and the average height is about 30 m. The terrain consists of three types, i.e., dry soil, wet soil, and vegetation. We randomly set eight RF transmitters, e.g., pedestrians, vehicles, micro base stations, and so on, and each one has a directional or isotropic antenna. The main simulation parameters are shown in the Table \ref{table1}. 

In the simulation, we firstly discretize the ROI into $N = 250 \times 250 \times 6 = 375000$ cubes and each cube is ${\text{5m}} \times 5{\text{m}} \times 10{\text{m}}{\text{.}}$ Since it is quite difficult to measure the spectrum data due to high-cost hardware system and stability of spectrum situation, we leverage the RT technique to calculate the 3D REM for the performance evaluation purpose. The RT technique has been widely used for radio propagation modeling, and has good performance on the RSS prediction under the specific area. The calculated REM is denoted as the ideal REM by a spectrum tensor $\chi  \in {\Re ^{250 \times 250 \times 6}}$ as shown in Fig. \ref{fig3} (b). Note that the REM is typically divided into cubes and we obtain the median signal strength from multiple measurements in RT simulation to remove the effects of small-scale fading\cite{Yilmaz13CM}.

Seven construction methods including different sampling positions and different recovery methods are conducted via different sampling rates. The proposed hierarchical construction algorithm is denoted as SNLO-SBLHM, where SNLO represents the proposed sampling optimization algorithm in Algorithm \ref{alg1} and SBLHM denotes the hierarchical recovery method in Algorithm \ref{alg2}. Lasso denotes the popular CS-based least-absolute shrinkage and selection operator (Lasso)\cite{Bazerque10TSP}. The Kriging algorithm and the low-rank tensor (matrix) completion HALRTC algorithm \cite{Liu13TPAMIL}, are also included as the representation of data-driven methods.  Moreover, four sampling optimization schemes, i.e., SNLO, Random sampling, the determinant-based greedy (DG) sampling algorithm \cite{Shen22WCL}, and FrameSense sampling algorithm\cite{Ranieri14TSP}, are also considered. The channel propagation model is defined as the free-space path loss model with the path loss exponent $\eta $ of 2 \cite{Bazerque10TSP}. The hyperparameters are generally given fixed small values to make these priors non-information, i.e., $a_0 = b_0 = 10^{-6}$, $c_0 = d_0 = 0$.

\begin{table}[htbp]
	\centering
	\caption{The main simulation parameters}
	\label{table1} 
	\renewcommand\arraystretch{1.4}
	\begin{tabular}{|c|clccc|}
		\hline
		\multirow{2}{*}{Parameter}          & \multicolumn{5}{c|}{Value}                                                                                                                                  \\ \cline{2-6} 
		& \multicolumn{2}{c|}{Index} & \multicolumn{1}{c|}{Height(m)} & \multicolumn{1}{c|}{\begin{tabular}[c]{@{}c@{}}Power\\    (dBmW)\end{tabular}} & Antenna Type \\ \hline
		\multirow{8}{*}{RF emitter setting} & \multicolumn{2}{c|}{1}     & \multicolumn{1}{c|}{1.5}       & \multicolumn{1}{c|}{20}                                                        & Isotropic    \\ \cline{2-6} 
		& \multicolumn{2}{c|}{2}     & \multicolumn{1}{c|}{1.5}       & \multicolumn{1}{c|}{20}                                                        & Isotropic    \\ \cline{2-6} 
		& \multicolumn{2}{c|}{3}     & \multicolumn{1}{c|}{1.5}       & \multicolumn{1}{c|}{20}                                                        & Isotropic    \\ \cline{2-6} 
		& \multicolumn{2}{c|}{4}     & \multicolumn{1}{c|}{1}         & \multicolumn{1}{c|}{20}                                                        & Isotropic    \\ \cline{2-6} 
		& \multicolumn{2}{c|}{5}     & \multicolumn{1}{c|}{1}         & \multicolumn{1}{c|}{20}                                                        & Isotropic    \\ \cline{2-6} 
		& \multicolumn{2}{c|}{6}     & \multicolumn{1}{c|}{20}        & \multicolumn{1}{c|}{20}                                                        & Directional  \\ \cline{2-6} 
		& \multicolumn{2}{c|}{7}     & \multicolumn{1}{c|}{30}        & \multicolumn{1}{c|}{20}                                                        & Isotropic    \\ \cline{2-6} 
		& \multicolumn{2}{c|}{8}     & \multicolumn{1}{c|}{2}         & \multicolumn{1}{c|}{20}                                                        & Isotropic    \\ \hline
		Center frequency                    & \multicolumn{5}{c|}{2.45GHz}                                                                                                                                \\ \hline
		Cartography area                    & \multicolumn{5}{c|}{$1250{\text{m}} \times 1250{\text{m}} \times 50{\text{m}}{\text{.}}$}                                                                   \\ \hline
		Grid resolution                     & \multicolumn{5}{c|}{${\text{5m}} \times 5{\text{m}} \times 10{\text{m}}{\text{.}}$}                                                                         \\ \hline
		REM tensor size                     & \multicolumn{5}{c|}{${N_x} \times {N_y} \times {N_z} = 250 \times 250 \times 6$}                                                                            \\ \hline
	\end{tabular}
\end{table}

\begin{figure*}[!t]
	\centering
	\includegraphics[width=5in]{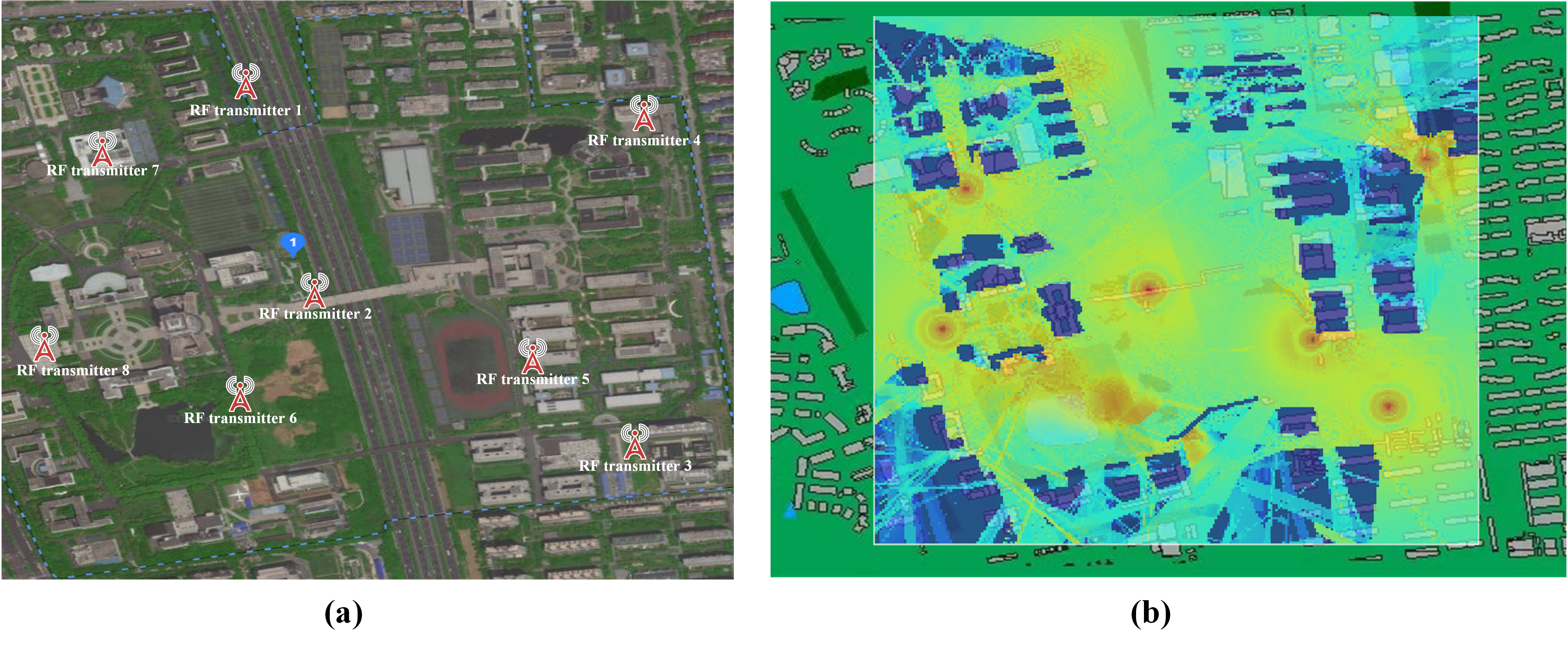}
	\caption{The satellite map and ideal REM (h = 2m) of campus scenario.}
	\label{fig3}
\end{figure*}

\subsection{3D REM Construction Performance}
To demonstrate the performance of different methods, we define the mean absolute error (MAE) as the averaged difference in RSS values between the reconstructed REM and ideal REM. It can be expressed as
\begin{equation}
MA{E^{{\text{REM}}}}{\text{ = }}\frac{1}{N}\sum\limits_{i = 1}^N {\left| {P_{^{{\text{est}}}}^r(i) - P_{{\text{true}}}^r(i)} \right|} ,
\label{eq63}
\end{equation}
where $P_{^{{\text{est}}}}^r(i)$ and $P_{^{{\text{true}}}}^r(i)$ are the estimated and true RSS in dBm at the $i\text{th}$ cube, respectively. 

As we can see from Fig. \ref{fig4}, the MAEs of SNLO-SBLHM, DG-SBLHM, FrameSense-SBLHM, Random-SBLHM, Random-Lasso, Random-HALRTC, and Random-Kriging decrease as the sampling rate increases. The proposed SNLO-SBLHM consistently outperforms other methods even at very low sampling rates. Besides, compared with the Random-SBLHM and Random-Lasso, the proposed hierarchical method improves the recovery performance substantially by considering the shadow fading under realistic scenarios. The proposed method outperforms other data-driven methods, i.e., Kriging and HALRTC, with a performance improvement of at least 50\%. Additionally, compared to Random-SBLHM and Random-Lasso, the proposed method substantially improves the recovery performance by taking into account shadow fading in realistic scenarios. Notably, when compared to Lasso, the proposed method shows an improvement of about 7dB.

On the other hand, it is evident that sampling location selection is vital for the REM construction, particularly when involving limited sampling data. Two data-driven algorithms' performances are unsatisfactory at low sampling rates since they usually need sufficient data to extract the spectrum correlations.  

Based on Fig. \ref{fig11}, it is clear that the DG-SBLHM algorithm consumes much more computational resources than  SNLO-SBLHM, which shows the complexity reduction works, i.e., the PCA preprocessing and the proposed efficient sampling scheme. However, Fig. \ref{fig4} shows that the performances of SNLO-SBLHM and DG-SBLHM are comparable as $r$ increases, which also demonstrates that the PCA preprocessing can effectively maintain the performance with appropriately retaining essential features. Moreover, Random-SBLHM, Random-Lasso, Random-HALRTC, and Random-Kriging present comparable computational efficiency and are much faster than DG-SBLHM, FrameSense-SBLHM and SNLO-SBLHM, as their sampling locations are randomly generated without optimization. Besides, although FrameSense-SBLHM consumes less time than SNLO-SBLHM, the SNLO-SBLHM shows greater advantages in construction performance, which is reasonable.
\begin{figure}[!t]
	\centering
	\includegraphics[width=3.5in]{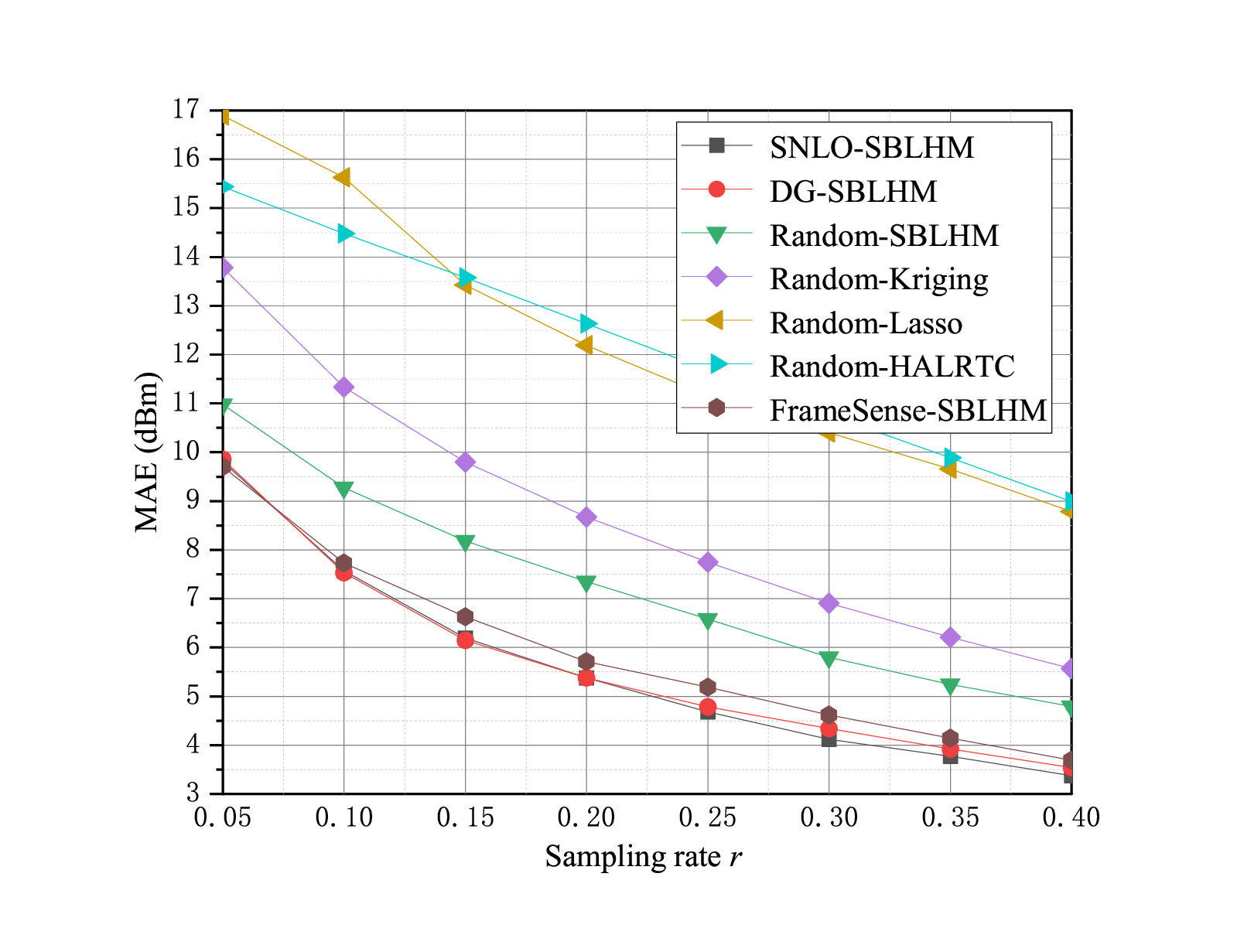}
	\caption{3D REM construction performance comparisons.}
	\label{fig4}
\end{figure}

\begin{figure}[!t]
	\centering
	\includegraphics[width=3.5in]{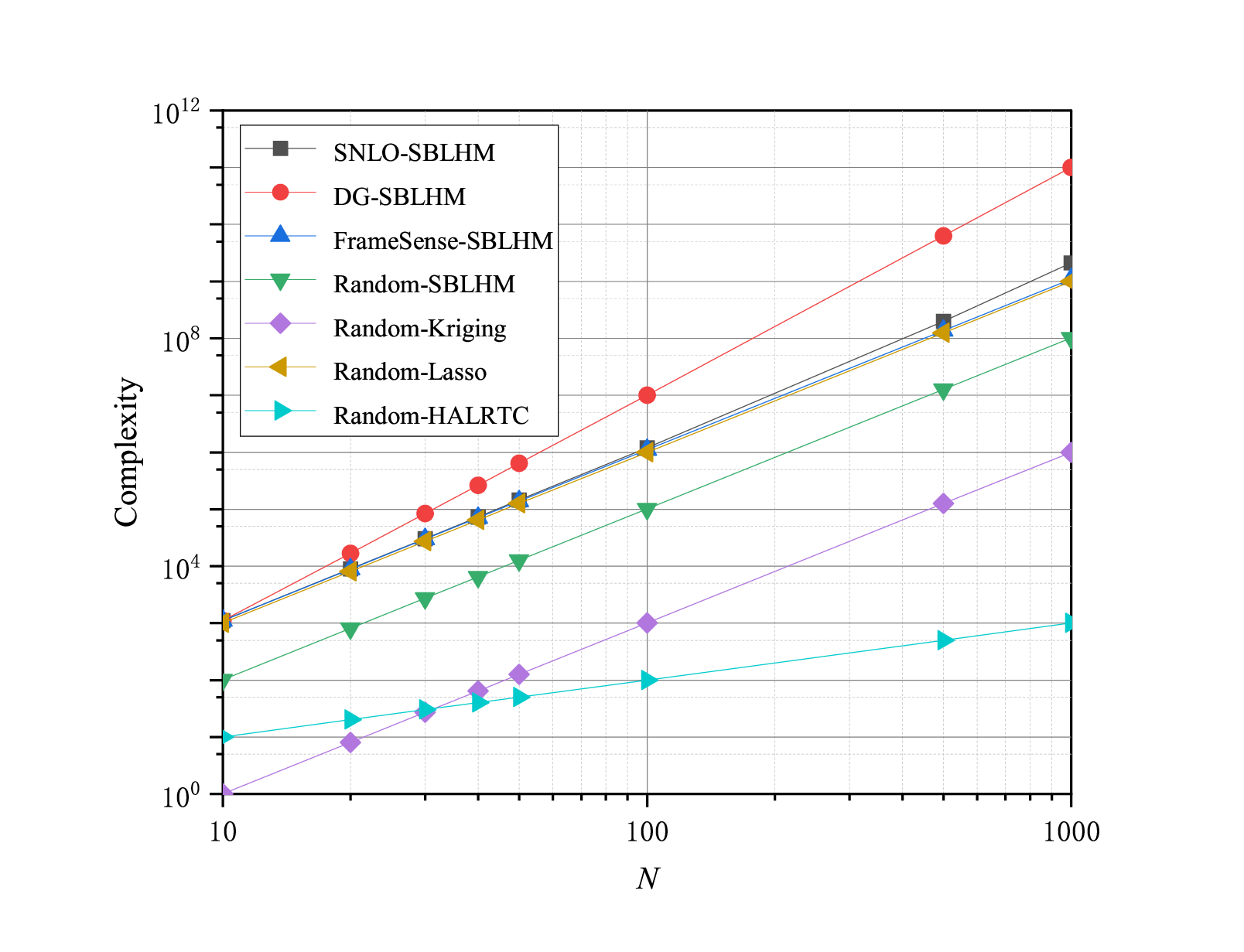}
	\caption{Performance comparisons of complexity.}
	\label{fig11}
\end{figure}
\subsection{Sampling Optimization Performance}
The WC-SBL-V index of different sampling schemes and the corresponding MAEs are given in Fig. \ref{fig6}. Four sampling schemes are compared in this paper, i.e., SNLO, DG, FrameSense, and random. It is shown that if the sampling number is the same, the SNLO provides the best performance or the minimum WC-SBL-V. In other words, if we fix the WC-SBL-V index threshold, the proposed SNLO algorithm requires the least number of sampling number to meet the accuracy requirement. It is more prominent when the sampling rate is low. In Fig. \ref{fig6}, it is clearly shown that SNLO outperforms the FrameSense and random sampling in identifying critical sensing locations, especially when the number of samples is limited. Besides, the SNLO and DG have comparable performances. DG considers the influence of all eigenvalues’ product, i.e., determinants, whereas SNLO only considers the influence of the largest eigenvalue. FrameSense considers the sum of all eigenvalues. While the maximum eigenvalue contributes the main part of determinant, especially when the sampling rate is low \cite{Jiang16TSP}. This also provides an explanation for why DG and SNLO outperform FrameSense. Therefore, it is sufficient to focus on the influence of maximum eigenvalue, i.e., WC-SBL-V index, to achieve satisfactory performance. Moreover, SNLO can greatly reduce the computational complexity than DG. It is apparent from Fig. \ref{fig6} that random sampling yields the worst result, reaffirming the crucial importance of spatial sampling location selection.

\begin{figure}[!t]
	\centering
	\includegraphics[width=3.5in]{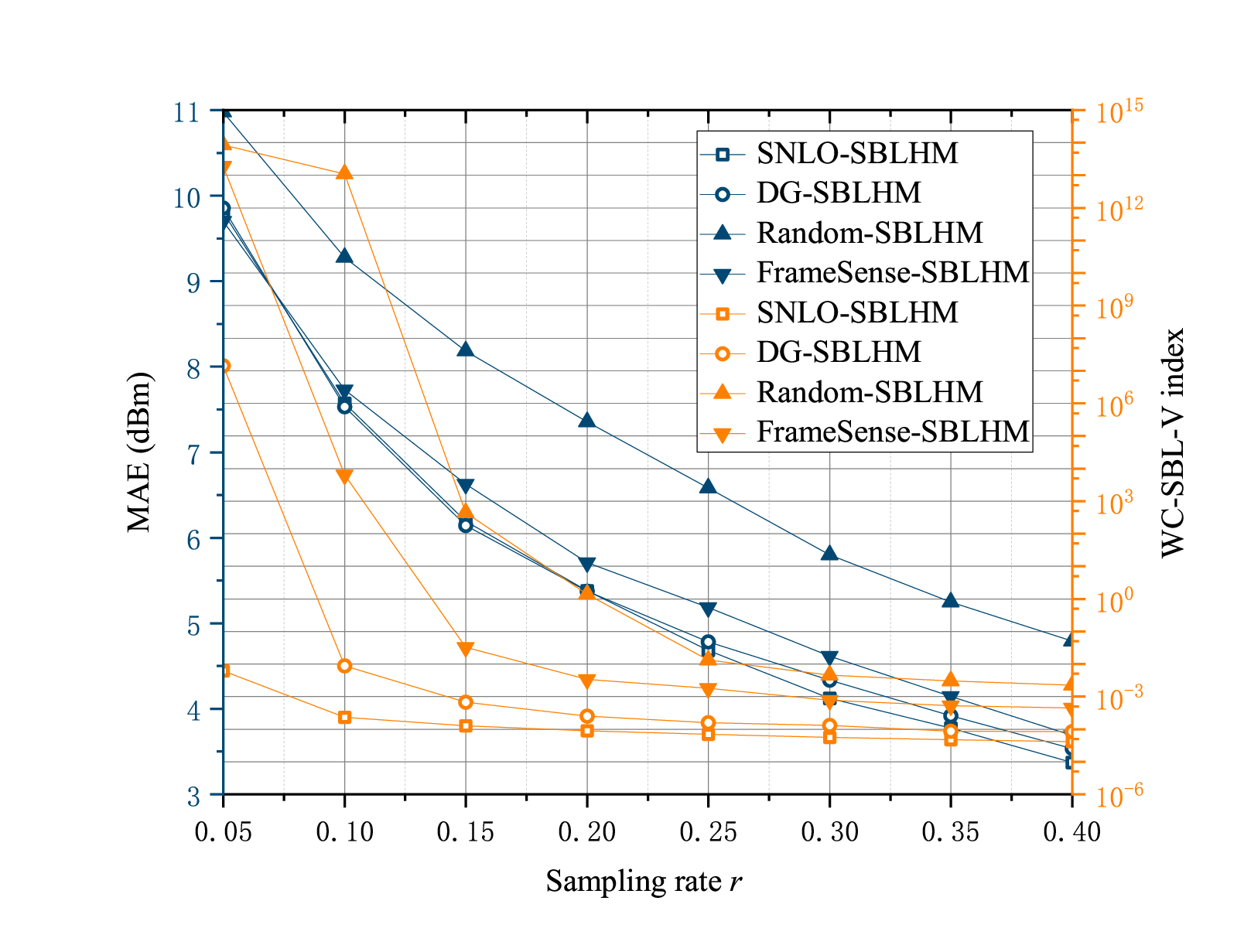}
	\caption{Comparison of sampling optimization performance.}
	\label{fig6}
\end{figure}

\subsection{Impact of Different Parameters}
The impact of the proportion threshold on the REM construction performance and WC-SBL-V index is shown in Figs. \ref{fig7} and \ref{fig8}, respectively. It can be seen that the performance of proposed algorithm generally improves with the decrease of $Per\_thr$. The WC-SBL-V index also decreases as the decrease of proportion threshold, which aligns with the objective function when optimizing sampling. In Fig.\ref{fig8}, it can be observed that the MAE shows minimal fluctuations with different $Per\_thr$. This indicates that the algorithm's performance is not significantly affected by slight errors. However, the threshold has a considerable impact on the complexity of the algorithm. Accordingly, by retaining more principal components of sparse dictionary, we can reduce the sparse signal recovery errors, and thus obtain an improved spectrum data recovery performance. However, with the increase of the proportion of retained principal components, the corresponding matrix dimension also increases, which results in an increase in computational complexity. Then, we need to reasonably balance the computational complexity and the performance, especially for the case of high-dimensional data. Note that when the matrix is high-dimensional, the selected sampling locations are also affected, which can result in the final selection being too close with each other and bring unsatisfactory recovery performance.

The impact of SNR on the construction performance is shown in Fig. \ref{fig9}. The SBL-based algorithms have better performance than others which demonstrates that the SBL-based method has an excellent noise supression capability. Based on this feature, the proposed SNLO-SBLHM further improves the recovery accuracy by optimizing the sampling positions and considering the shadow fading. In addition, it can be observed that the Kriging algorithm has the worst performance in the presence of noise. The reason is that it relies on the geographical location and measured data to estimate the spectrum data at the unsampled positions, which cannot identify the noise influence.

\begin{figure}[!t]
	\centering
	\includegraphics[width=3.5in]{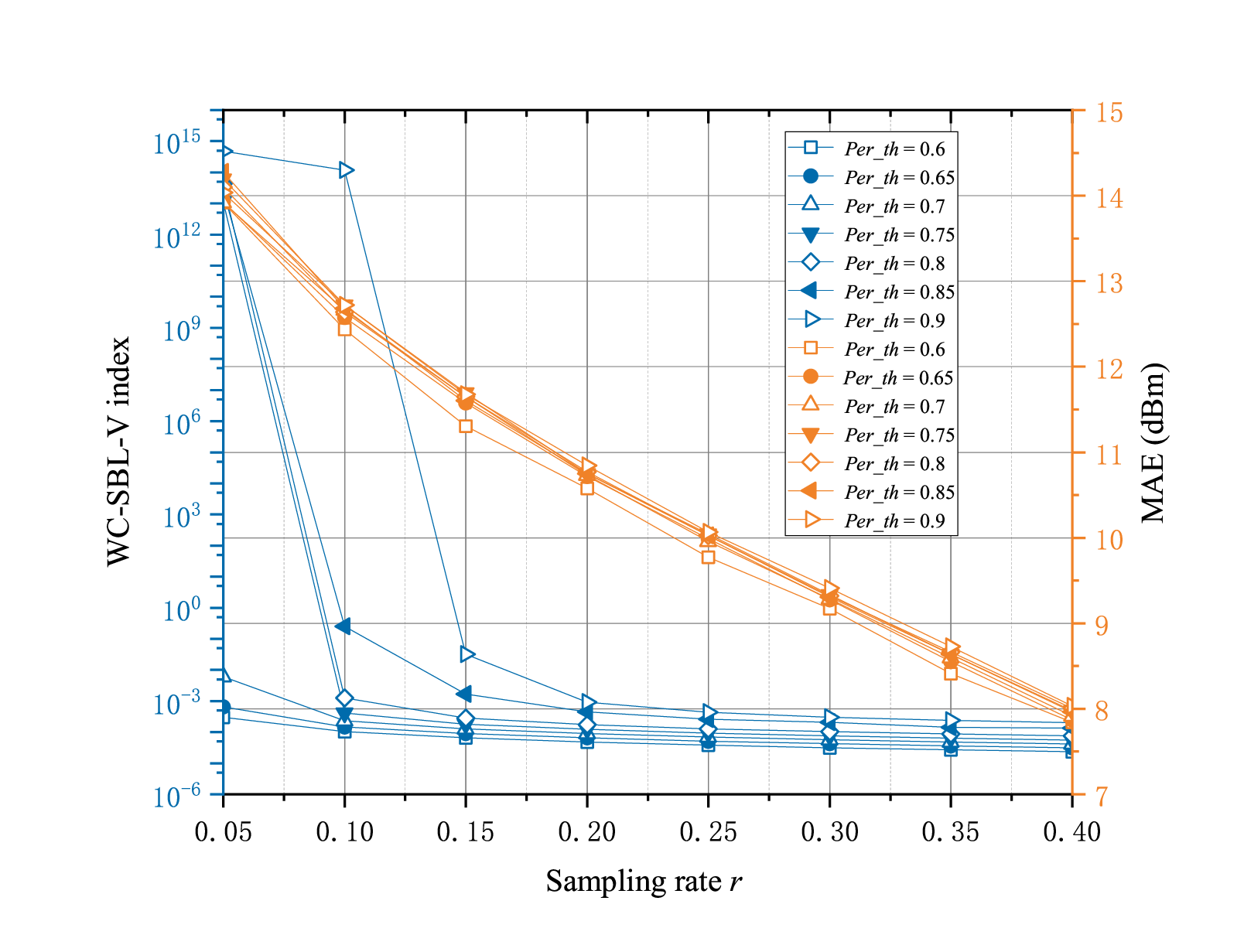}
	\caption{Impact of $Per\_thr$ on the construction performances and the WC-SBL-V index.}
	\label{fig7}
\end{figure}

\begin{figure}[!t]
	\centering
	\includegraphics[width=3.5in]{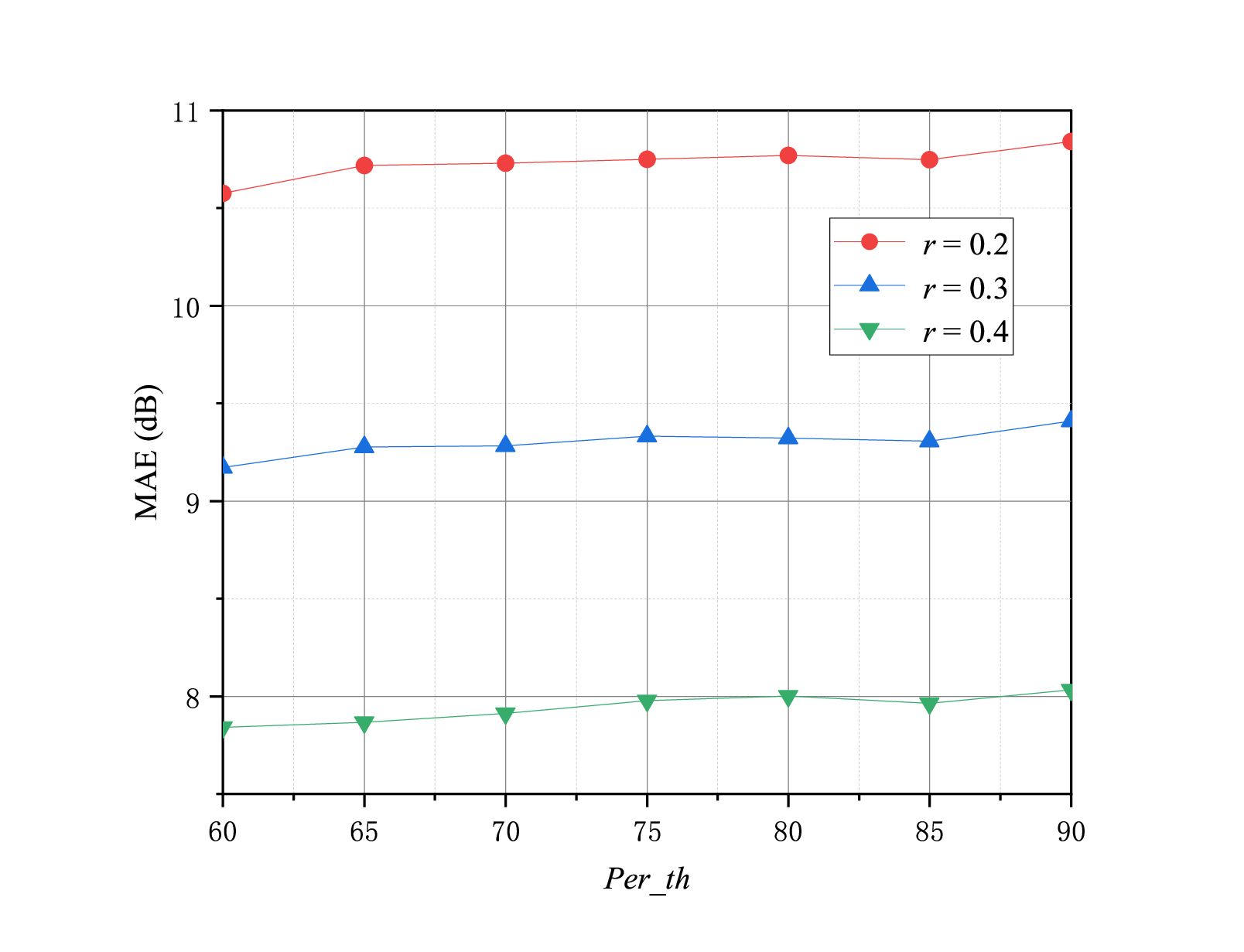}
	\caption{Impact of $Per\_thr$ on the construction performance.}
	\label{fig8}
\end{figure}

\begin{figure}[!t]
	\centering
	\vspace{-0.5cm}
	\includegraphics[width=3.5in]{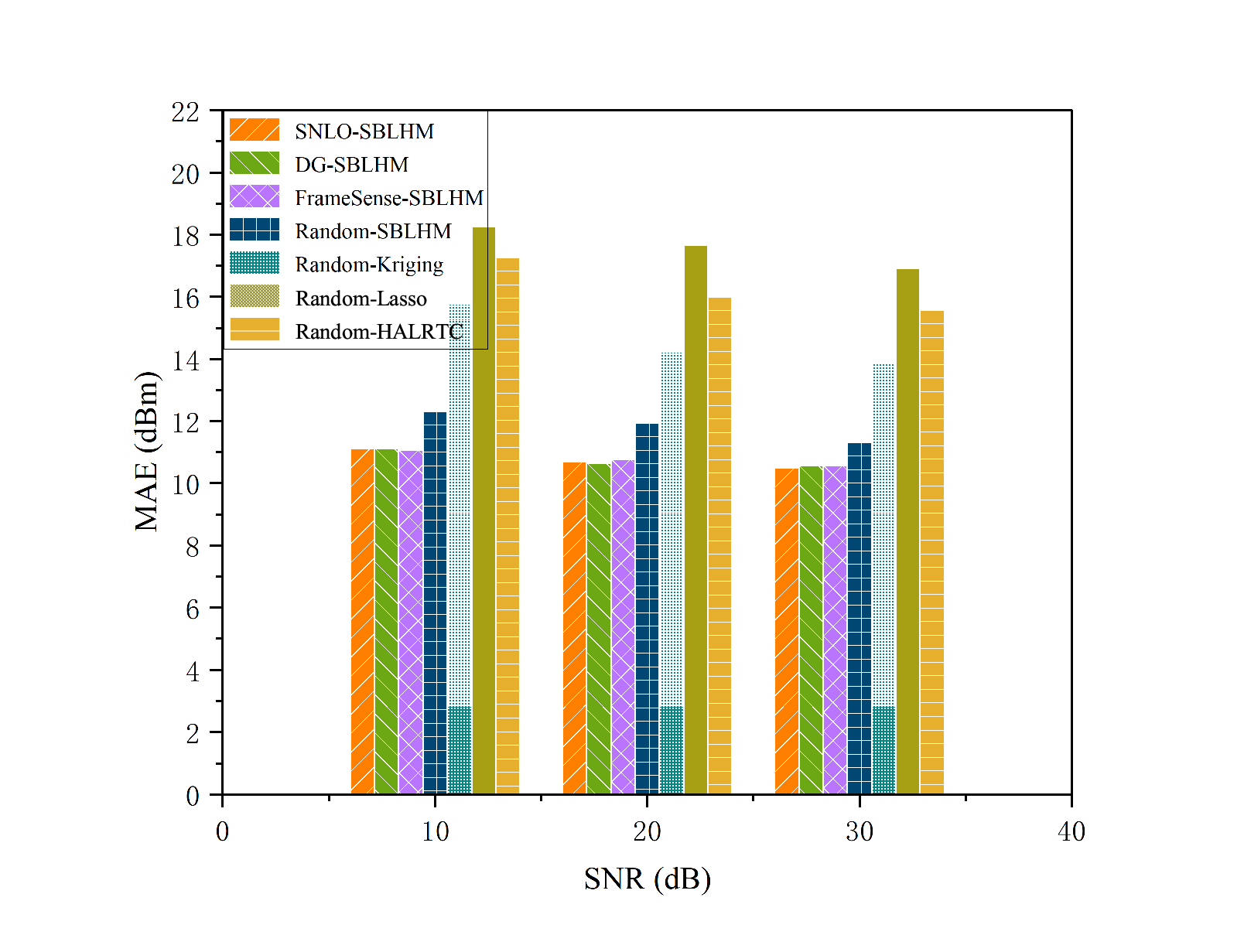}
	\caption{Impact of SNR on the construction performance ($r$ = 0.05).}
	\label{fig9}
\end{figure}

\begin{figure*}[!t]
	\centering
	\includegraphics[width=7in]{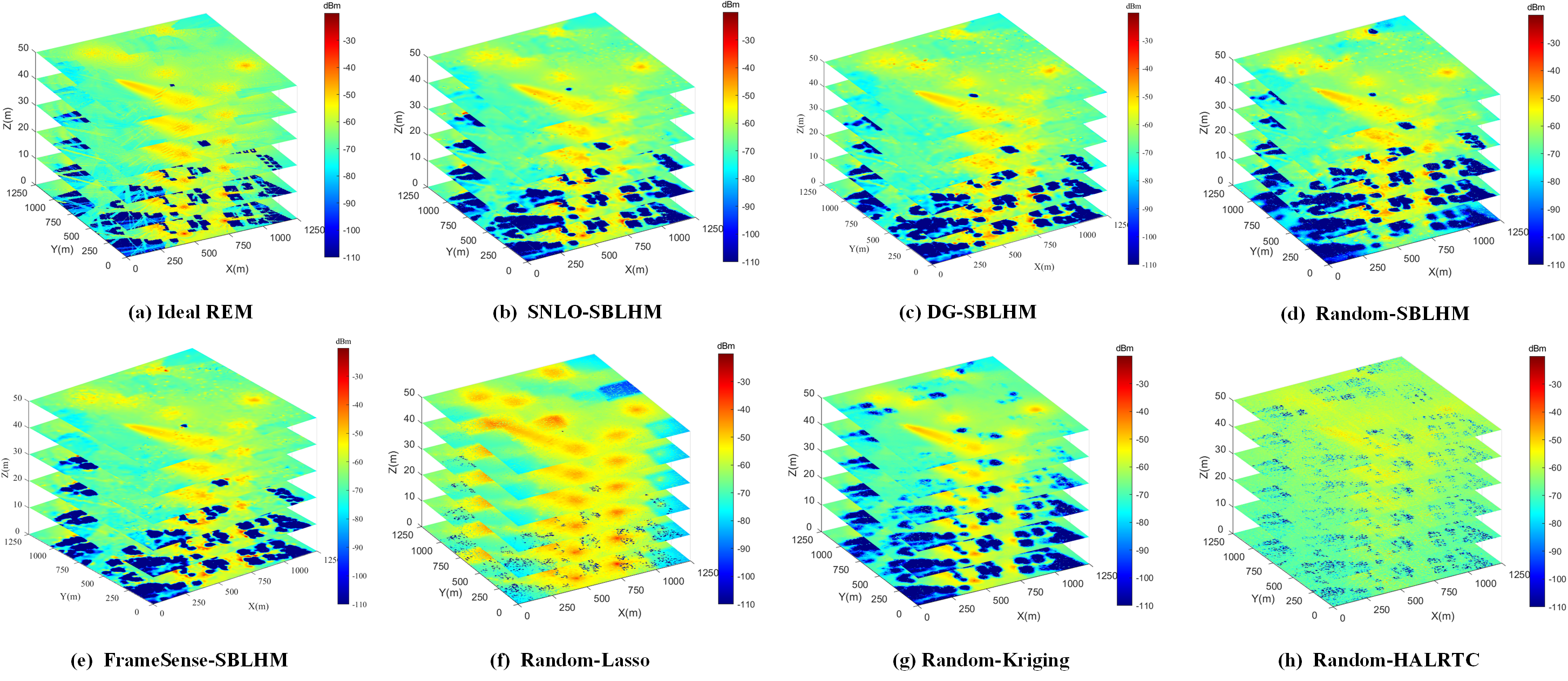}
	\caption{Visualization of constructed 3D REMs with different algorithms ($r$ = 0.1). (a) Ideal REM; and (b)-(h) SNLO-SBLHM, DG-SBLHM, Random-SBLHM, FrameSense-SBLHM, Random-Lasso, Random-Kriging, and Random-HALRTC, respectively.}
	\label{fig10}
\end{figure*}

\subsection{REM Construction Performance Visualization}
In order to intuitively present the construction performance of different methods under realistic scenarios, Fig. \ref{fig10} compares the constructed 3D REMs at different altitudes under the campus scenario. It can be seen that the SNLO-SBLHM can achieve the most effective spectrum situation recovery. This is especially noticeable for the spectrum data near the RF transmitters or around the buildings. Fig. \ref{fig10} (b)-(e) demonstrate that the recovery algorithms based on SBL exhibit superior performances in radiation source recovery. By comparing (g) and (h), it reveals that Kriging has better performance than HALRTC. This is because that HALRTC requires the low-rank characteristic of spectrum tensor, which is relatively poor when there are a substantial number of RF transmitters or buildings.

\section{Conclusion}
This paper has addressed the issue of 3D REM construction with limited sample positions under unknown environments. We have formulated the 3D REM construction as the sparse sampling optimization and data recovery problems, by exploiting the sparsity of spectrum data and the characteristics of channel propagation. Then, a worst case SBL variance-based measurement matrix optimization mechanism has been developed to determine the required sampling quantity and sampling locations under the given recovery threshold. Meanwhile, PCA preprocessing has been performed on the sparse dictionary to further improve the sampling efficiency by reducing the matrix dimension. We have also proposed a hierarchical SBL model for the spectrum data recovery. The sparse signal has been recovered by SBL in the first layer and the shadow fading components have been estimated by GPR in the second layer. The construction performance of typical algorithms, i.e., SNLO-SBLHM, DG-SBLHM, FrameSense-SBLHM, Random-SBLHM, Random-Lasso, Random-HALRTC, and Random-Kriging have been fully compared and analyzed. The impacts of sparsity and proportion threshold on the recovery accuracy have also been studied. It has been shown that the proposed sampling and recovery methods are effective under unknown scenarios and superior in 3D REM construction against other methods. In the future work, we will apply the hierarchical recovery method on the UAV-assisted 3D REM mapping system \cite{DEMO22, Mao23TIM} and optimize the UAV trajectory based on the proposed sampling scheme.

\appendices
\section{Derivation of equations (\ref{eq19}) and (\ref{eq20})}
\label{append1x b}
To obtain (\ref{eq19}) and (\ref{eq20}), we need to compute marginal likelihood $p\left( {\bm{t}|\bm{\alpha} ,\beta } \right)$ term in (18). We first derive the term ${\mathbf{\Lambda }}$ by exploiting the determinant identity
\begin{equation}
\left| \mathcal{A} \right|\left| {\beta _{}^{ - 1}{\mathbf{I}} + {\mathbf{\Phi }}\mathcal{A}_{}^{ - 1}{{\mathbf{\Phi }}^{\text{T}}}} \right| = \left| {\beta _{}^{ - 1}{\mathbf{I}}} \right|\left| {\mathcal{A} + \beta {{\mathbf{\Phi }}^{\text{T}}}{\mathbf{\Phi }}} \right|.
\label{eq69}
\end{equation}
We have
\begin{equation}
\begin{aligned}
\left| {\mathbf{\Lambda }} \right|_{}^{ - 1/2} &= \left| {\beta _{}^{ - 1}{\mathbf{I}} + {\mathbf{\Phi }}\mathcal{A}_{}^{ - 1}{{\mathbf{\Phi }}^{\text{T}}}} \right|_{}^{ - 1/2}, \\ 
&= \prod\limits_M {\beta _{}^{1/2}} \left| \mathcal{A} \right|_{}^{1/2}\left| {\mathcal{A} + \beta {{\mathbf{\Phi }}^{\text{T}}}{\mathbf{\Phi }}} \right|_{}^{ - 1/2}, \\ 
&= \beta _{}^{M/2}\prod\limits_{i = 1}^N {\alpha _i^{1/2}} \left| {\mathcal{A} + \beta {{\mathbf{\Phi }}^{\text{T}}}{\mathbf{\Phi }}} \right|_{}^{ - 1/2}. \\ 
\end{aligned} 
\label{eq70}
\end{equation}
Then, the Woodbury inversion identity is employed to the term $\left( {\mathbf{\Lambda }} \right)_{}^{ - 1}$ as
\begin{equation}
\begin{aligned}
\left( {\mathbf{\Lambda }} \right)_{}^{ - 1} &= \left( {\beta _{}^{ - 1}{\mathbf{I}} + {\mathbf{\Phi }}\mathcal{A}_{}^{ - 1}{{\mathbf{\Phi }}^{\text{T}}}} \right)_{}^{ - 1}, \\ 
&= \beta {\mathbf{I}} - \beta {\mathbf{\Phi }}\left( {\beta {{\mathbf{\Phi }}^{\text{T}}}{\mathbf{\Phi }} + \mathcal{A}} \right)_{}^{ - 1}{{\mathbf{\Phi }}^{\text{T}}}\beta , \\ 
\end{aligned} 
\label{eq71}
\end{equation}
we have
\begin{equation}
\begin{aligned}
t_{}^{\text{T}}\left( {\mathbf{\Lambda }} \right)_{}^{ - 1}t &= t_{}^{\text{T}}\left( {\beta {\mathbf{I}} - \beta {\mathbf{\Phi }}\left( {\beta {{\mathbf{\Phi }}^{\text{T}}}{\mathbf{\Phi }} + \mathcal{A}} \right)_{}^{ - 1}{{\mathbf{\Phi }}^{\text{T}}}\beta } \right)t, \\ 
&= \beta t_{}^{\text{T}}t - \beta t_{}^{\text{T}}{\mathbf{\Phi }}\left( {\beta {{\mathbf{\Phi }}^{\text{T}}}{\mathbf{\Phi }} + \mathcal{A}} \right)_{}^{ - 1}{{\mathbf{\Phi }}^{\text{T}}}\beta t, \\ 
&= \beta t_{}^{\text{T}}t - \beta t_{}^{\text{T}}{\mathbf{\Phi }}{{\mathbf{\Sigma }}_\omega }{{\mathbf{\Phi }}^{\text{T}}}\beta t, \\ 
&= \beta t_{}^{\text{T}}\left( {t - {\mathbf{\Phi }}\mu } \right), \\ 
\end{aligned} 
\label{eq72}
\end{equation}
where we define
\begin{equation}
\begin{gathered}
\bm{\mu}  = \beta {{\mathbf{\Sigma }}_\omega }{{\mathbf{\Phi }}^{\text{T}}}{\bm{t}}, \hfill \\
{{\mathbf{\Sigma }}_\omega } = {\left( {\beta {{\mathbf{\Phi }}^{\text{T}}}{\mathbf{\Phi }} + \mathcal{A}} \right)^{ - 1}}. \hfill \\ 
\end{gathered} 
\label{eq73}
\end{equation}
With (\ref{eq14}), (\ref{eq70}), and (\ref{eq73}), the marginal likelihood $p\left( {\bm{t}|\bm{\alpha} ,\beta } \right)$ is 
\begin{equation}
\begin{aligned}
& p\left( {\bm{t}|\bm{\alpha} ,\beta } \right) = \int {p\left( {\bm{t}|\bm{\omega} ,\beta } \right)p\left( {\bm{\omega} |\bm{\alpha} } \right)d\bm{\omega} } , \\ 
&= \left( {2\pi } \right)_{}^{ - M/2}\beta _{}^{M/2}\prod\limits_{i = 1}^N {\alpha _i^{1/2}} \left| {{{\mathbf{\Sigma }}_\omega }} \right|_{}^{1/2}\exp \left\{ { - \frac{1}{2}\beta \bm{t}_{}^{\text{T}}\left( {\bm{t} - {\mathbf{\Phi }}\bm{\mu} } \right)} \right\}, \\ 
\end{aligned} 
\label{eq74}
\end{equation}
Then, the weight posterior of $\bm{\omega}$ in (\ref{eq18}) is
\begin{equation}
\begin{aligned}
&p\left( {\bm{\omega} |\bm{t},\bm{\alpha} ,\beta } \right) = \frac{{p\left( {\bm{t}|\bm{\omega} ,\bm{\alpha} ,\beta } \right)p\left( {\bm{\omega} |\bm{\alpha} } \right)}}{{p\left( {\bm{t}|\bm{\alpha} ,\beta } \right)}}, \\ 
&= \left( {2\pi } \right)_{}^{ - N/2}\left| {{{\mathbf{\Sigma }}_\omega }} \right|_{}^{ - 1/2}\exp \left\{ { - \frac{1}{2}\left( {\bm{\omega}  - \bm{\mu} } \right)_{}^{\text{T}}\left( {{{\mathbf{\Sigma }}_\omega }} \right)_{}^{ - 1}\left( {\bm{\omega}  - \bm{\mu} } \right)} \right\}, \\ 
&= \mathcal{N}\left( {\bm{\omega} |\bm{\mu} ,{{\mathbf{\Sigma }}_\omega }} \right), \\ 
\end{aligned} 
\label{eq75}
\end{equation}
where the mean and variance is equivalent to (\ref{eq19}) and (\ref{eq20}).

\section{Derivation of equations (\ref{eq44}) and (\ref{eq46})}
\label{append1x c}
The objective (\ref{eq43}) is equivalent to maximize the product of hyper-priors and marginal likelihood $p\left( {\bm{t}|\bm{\alpha} ,\beta } \right)$, namely the “evidence for the hyper-parameters”. It is also known as the type-II maximum likelihood method, as 
\begin{equation}
\begin{aligned}
p\left( {\bm{t}|\bm{\alpha} ,\beta } \right) &= \int {p\left( {\bm{t}|\bm{\omega} ,\beta } \right)p\left( {\bm{\omega} |\bm{\alpha} } \right)d\bm{\omega} } , \\ 
&= \left( {2\pi } \right)_{}^{ - M/2}\left| {\mathbf{\Lambda }} \right|_{}^{ - 1/2}\exp \left\{ { - \frac{1}{2}\bm{t}_{}^{\text{T}}\left( {\mathbf{\Lambda }} \right)_{}^{ - 1}\bm{t}} \right\}, \\ 
\end{aligned} 
\label{eq76}
\end{equation}
where ${\mathbf{\Lambda }} = \beta _{}^{ - 1}{\mathbf{I}} + {\mathbf{\Phi }}\mathcal{A}_{}^{ - 1}{{\mathbf{\Phi }}^{\text{T}}}$. By ignoring the irrelevant terms, we can obtain the objective function of the priors over hyper-parameters in the logarithmic case as
\begin{equation}
\begin{aligned}
\mathcal{L}\left( {\bm{\alpha} ,\beta } \right) &=  - \frac{1}{2}\left\{ {\log \left| {\mathbf{\Lambda }} \right| + \bm{t}_{}^{\text{T}}\left( {\mathbf{\Lambda }} \right)_{}^{ - 1}\bm{t}} \right\} \\ 
&+ \sum\limits_{i = 0}^N {\left( {a\log {\alpha _i} - b{\alpha _i}} \right)}  + c\log \beta  - d\beta . \\ 
\end{aligned} 
\label{eq77}
\end{equation}
Then, we calculate the derivative with respect to $\log \alpha _i^{}$ and let $\partial \mathcal{L}\left( {\bm{\alpha} ,\beta } \right)/\partial \log \alpha _i^{} = 0$ to find the re-estimation rule of $\alpha _i^{}$ to obtain (\ref{eq44}). Similarly, calculating the derivative of (\ref{eq77}) with respect to $\log \beta $ and let $\partial \mathcal{L}\left( {\bm{\alpha} ,\beta } \right)/\partial \log \beta  = 0$ to obtain the update rule of $\beta $ in (\ref{eq46}).

%

\bibliographystyle{IEEEtran}
\bibliography{myref}


\newpage

\begin{IEEEbiography}[{\includegraphics[width=1in,height=1.25in,clip,keepaspectratio]{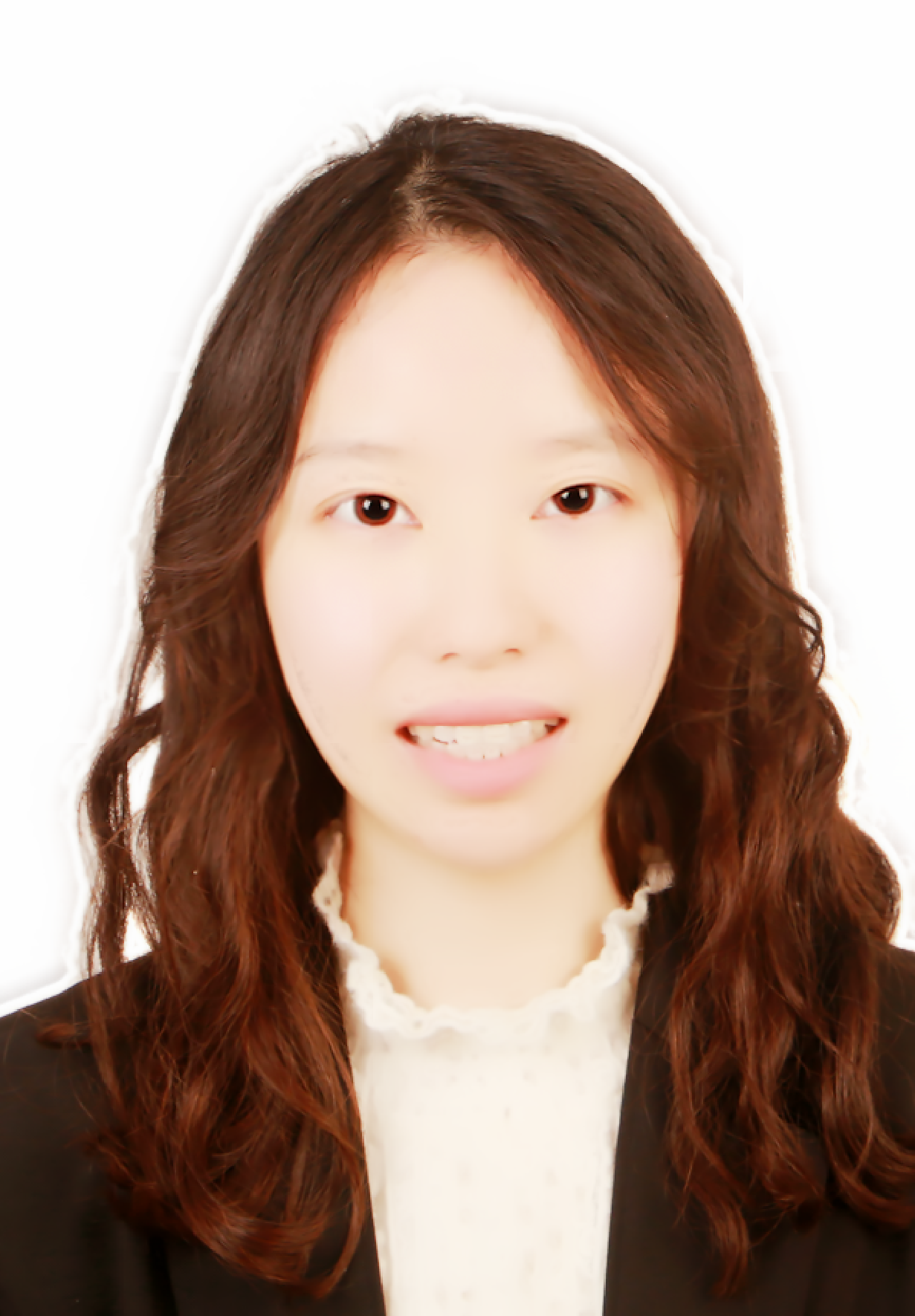}}]{Jie Wang}
received the B.S. degree in internet of things engineering from the College of Information Science and Technology, Nanjing Forestry University of China, Nanjing, China, in 2021. She is currently pursuing the Ph.D. degree in communications and information systems with the College of Electronic and Information Engineering, Nanjing University of Aeronautics and Astronautics. Her current research interests conclude spectrum mapping. 
\end{IEEEbiography}

\vspace{9pt}

\begin{IEEEbiography}[{\includegraphics[width=1in,height=1.25in,clip,keepaspectratio]{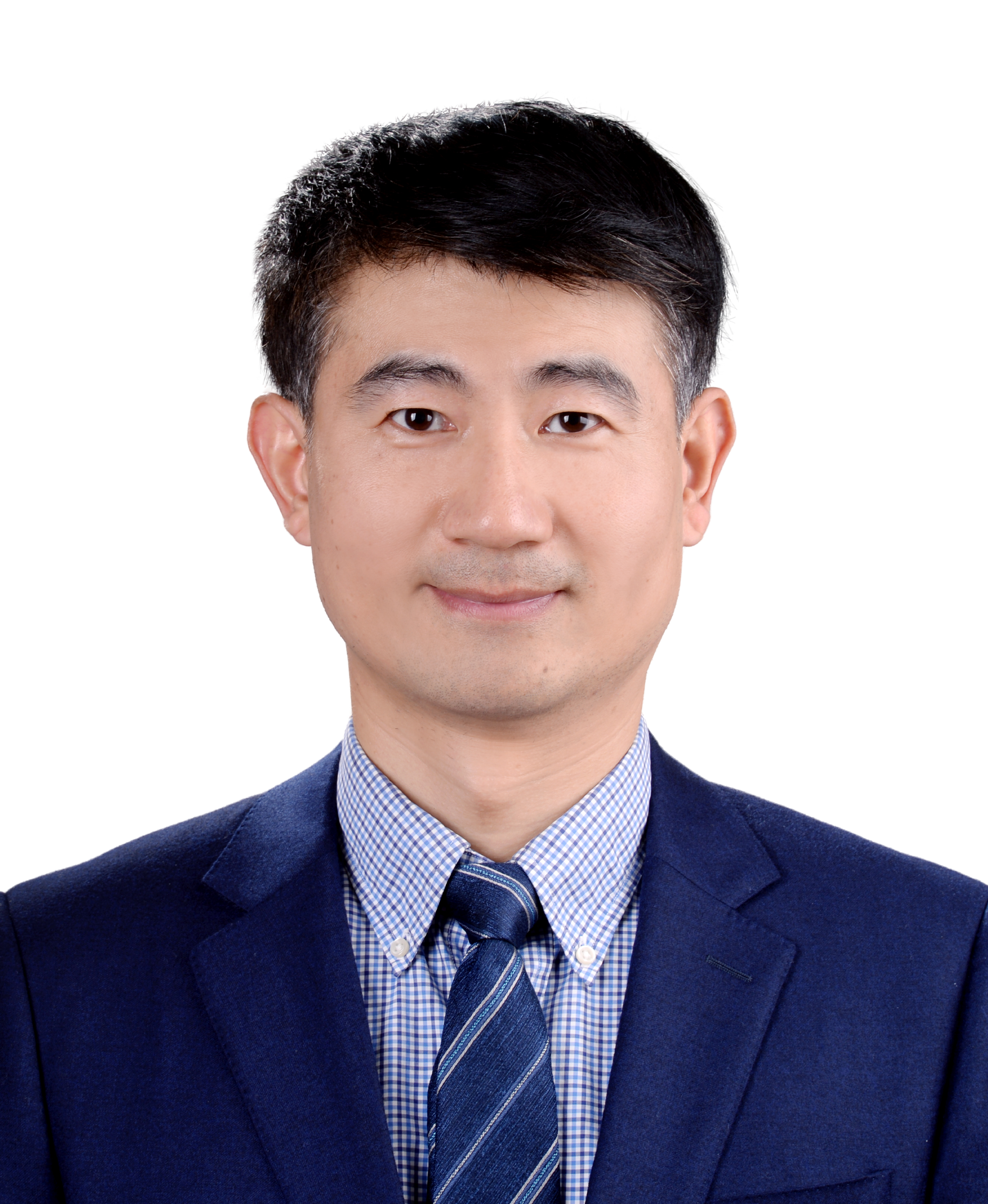}}]{Qiuming Zhu}
	received the B.S. degree in electronic engineering and the M.S. and Ph.D. degrees in communication and information system from the Nanjing University of Aeronautics and Astronautics (NUAA), Nanjing, China, in 2002, 2005, and 2012, respectively. He is currently a Professor in the College of Electronic and Information Engineering, Nanjing University of Aeronautics and Astronautics, Nanjing, China. His current research interests include channel sounding, modeling, and emulation for the fifth/sixth generation (5G/6G) mobile communication, vehicle-to-vehicle (V2V) communication, and unmanned aerial vehicles (UAV) communication systems.
\end{IEEEbiography}

\vspace{9pt}

\begin{IEEEbiography}[{\includegraphics[width=1in,height=1.25in,clip,keepaspectratio]{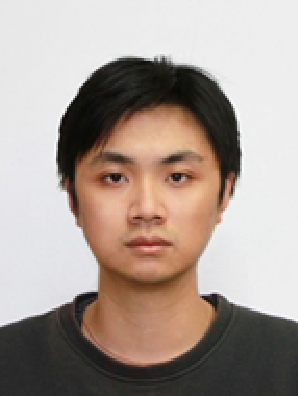}}]{Zhipeng Lin}
received the Ph.D. degrees from the School of Information and Communication Engineering, Beijing University of Posts and Telecommunications, Beijing, China, and the School of Electrical and Data Engineering, University of Technology of Sydney, NSW, Australia, in 2021. Currently, He is an Associate Researcher in the College of Electronic and Information Engineering, Nanjing University of Aeronautics and Astronautics, Nanjing, China. His current research interests include signal processing, massive MIMO, spectrum sensing, and UAV communications.
\end{IEEEbiography}

\vspace{9pt}

\begin{IEEEbiography}[{\includegraphics[width=1in,height=1.25in,clip,keepaspectratio]{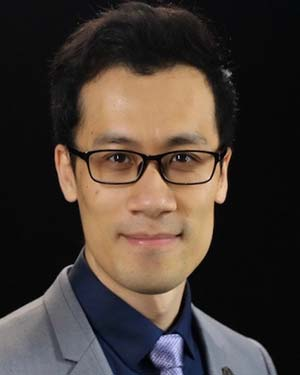}}]{Gunting Chen}
	(Member, IEEE) received the B.Sc. degree in electronic engineering from Nanjing University, Nanjing, China, in 2009, and the Ph.D. degree in electronic and computer engineering from the Hong Kong University of Science and Technology, Hong Kong, in 2015. From 2014 to 2015, he was a Visiting Student with the Wireless Information and Network Sciences Laboratory, Massachusetts Institute of Technology, Cambridge, MA, USA. He is currently an Assistant Professor with the School of Science and Engineering and the Future Network of Intelligence Institute (FNii), The Chinese University of Hong Kong, Shenzhen (CUHK–Shenzhen), Shenzhen, China. Prior to joining CUHK Shenzhen, he was a Postdoctora Research Associate with the Ming Hsieh Department of Electrical Engineering, University of Southern California, Los Angeles, CA, USA, from 2016 to 2018, and with the Communication Systems Department of EURECOM, Sophia–Antipolis, France, from 2015 to 2016. His research interests include channel estimation, MIMO beamforming, machine learning, and optimization for wireless communications and localization. His current research interests include radio map sensing, construction, and application for wireless communications. Dr. Chen was the recipient of the HKTIIT Post-Graduate Excellence Scholarships in 2012. He was nominated as the Exemplary Reviewer of IEEE W IRELESS C OMMUNICATIONS L ETTERS in 2018. His paper received the Charles Kao Best Paper Award from WOCC 2022.
\end{IEEEbiography}

\vspace{9pt}

\begin{IEEEbiography}[{\includegraphics[width=1in,height=1.25in,clip,keepaspectratio]{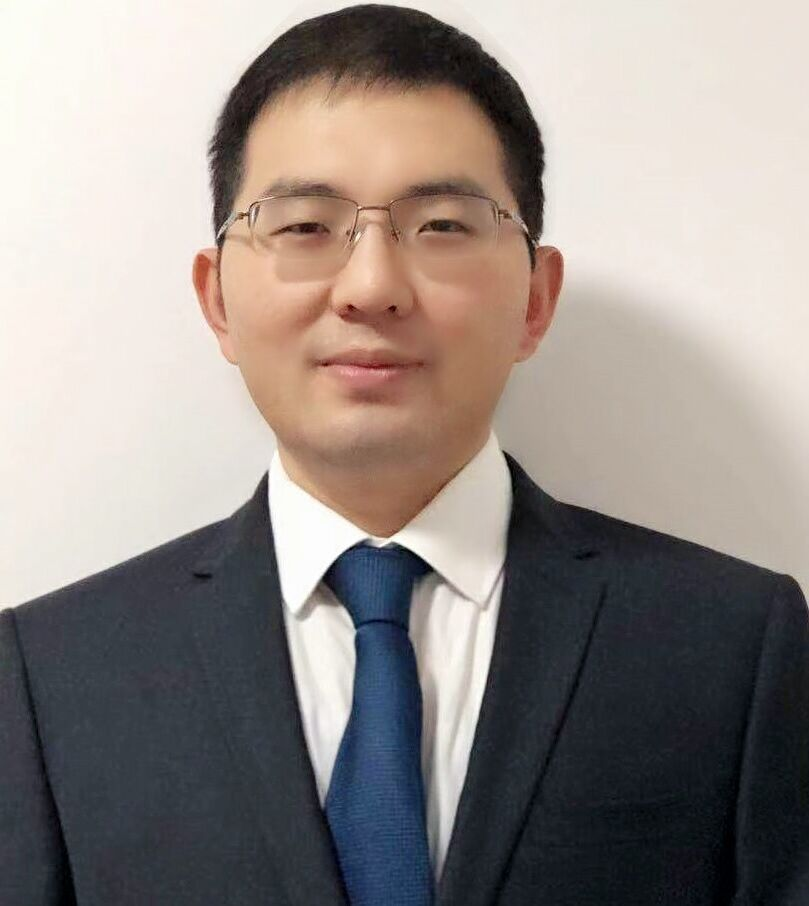}}]{Guoru Ding}
	(SM’16) received the B.S. (Hons.) degree in electrical engineering from Xidian University, Xi’an, China, in 2008, and the Ph.D. (Hons.) degree in communications and information systems from the College of Communications Engineering, Nanjing, China, in 2014. He is currently an associate professor in the College of Communications Engineering, Nanjing, China. His research interests include cognitive radio networks, massive MIMO, machine learning, and big data analytics over wireless networks. He is also an Associate Editor of IEEE Transactions on Cognitive Communications and Networking. He served as a Guest Editor for the IEEE Journal on Selected Areas in Communications Special Issue on Spectrum Sharing and Aggregation in Future Wireless Networks.
\end{IEEEbiography}

\vspace{9pt}

\begin{IEEEbiography}[{\includegraphics[width=1in,height=1.25in,clip,keepaspectratio]{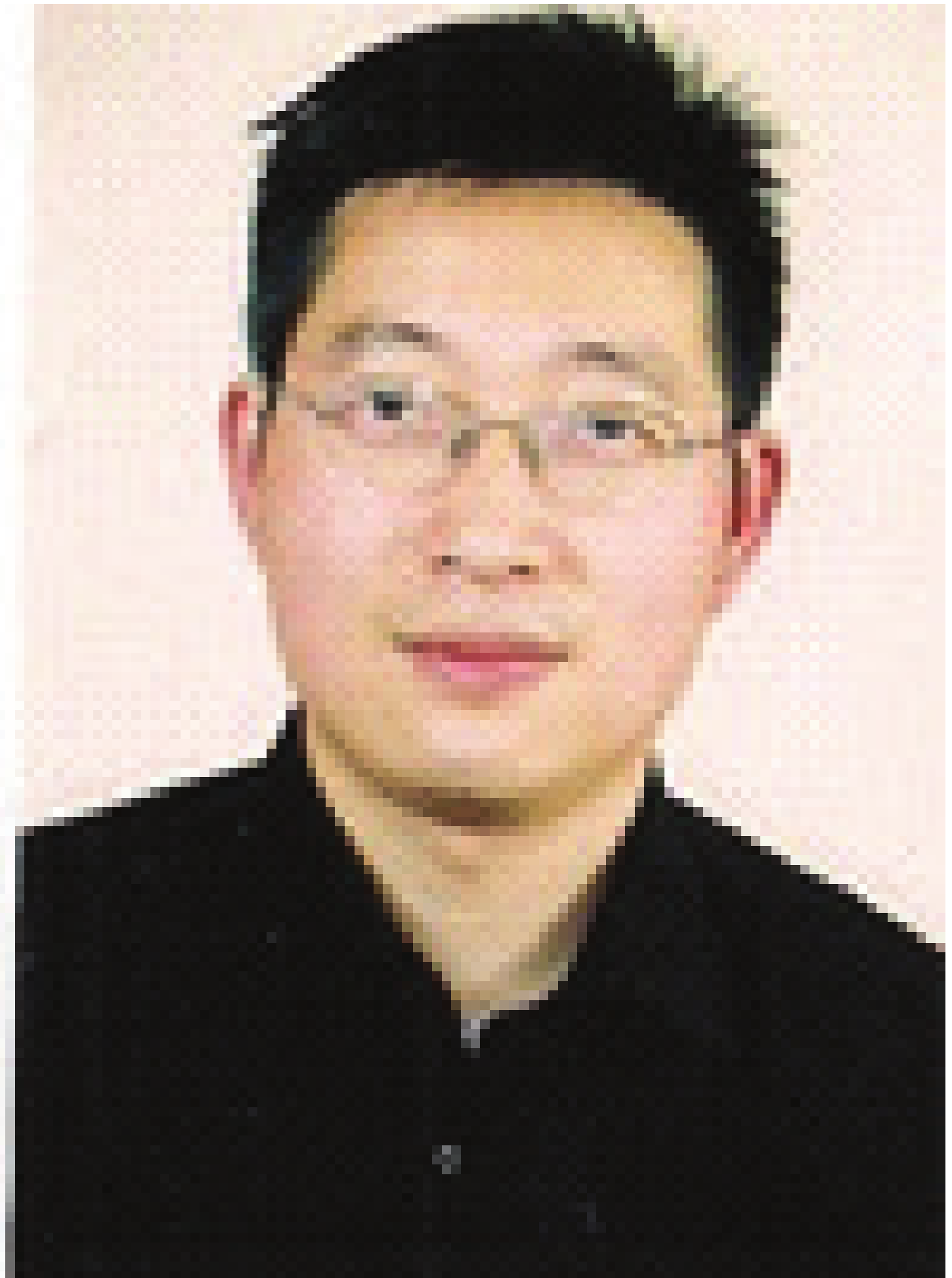}}]{Qihui Wu}
	received the B.S. degree in communications engineering and the M.S. and Ph.D. degrees in communications and information system from the PLA University of Science and Technology, Nanjing, China, in 1994, 1997, and 2000, respectively. He is currently a Professor with the College of Electronic and Information Engineering, Nanjing University of Aeronautics and Astronautics. His current research interests include algorithms and optimization for cognitive wireless networks, soft-defined radio, and wireless communication systems.
\end{IEEEbiography}

\vspace{9pt}

\begin{IEEEbiography}[{\includegraphics[width=1in,height=1.25in,clip,keepaspectratio]{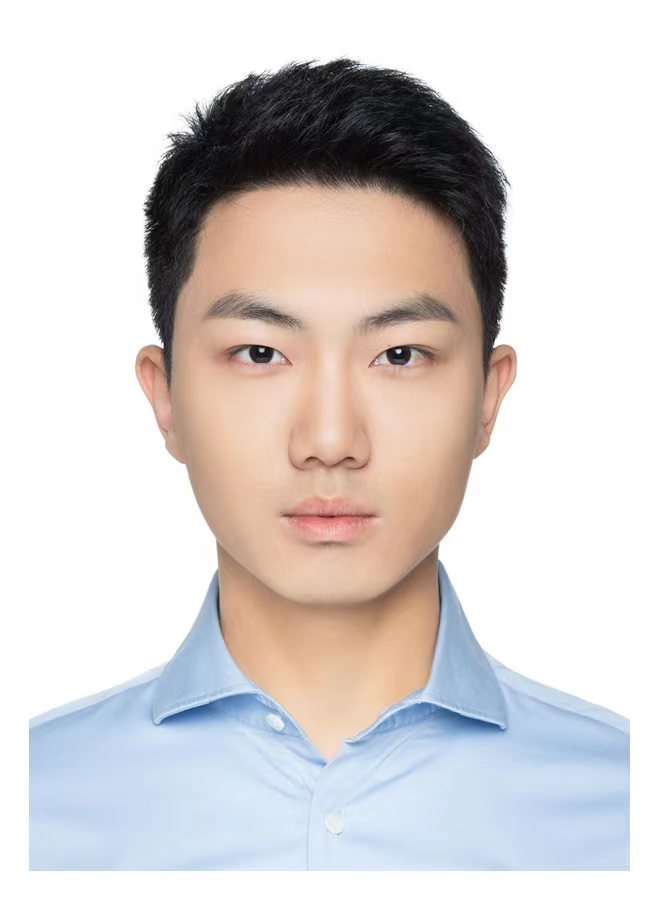}}]{Guochen Gu}
	received the B.S. degree in computer science from the School of the Gifited Young, University of Science and Technology of China in 2021. He received the M.S. degree in computer engineering from the School of Engineering, New York University in 2023. He is currently pursuing the P.H.D. degree in computer science in the College of Computer Science, Nanjing University of Aeronautics and Astronautics. His current rescarch direction is related to spectrum.
\end{IEEEbiography}

\vspace{9pt}

\begin{IEEEbiography}[{\includegraphics[width=1in,height=1.25in,clip,keepaspectratio]{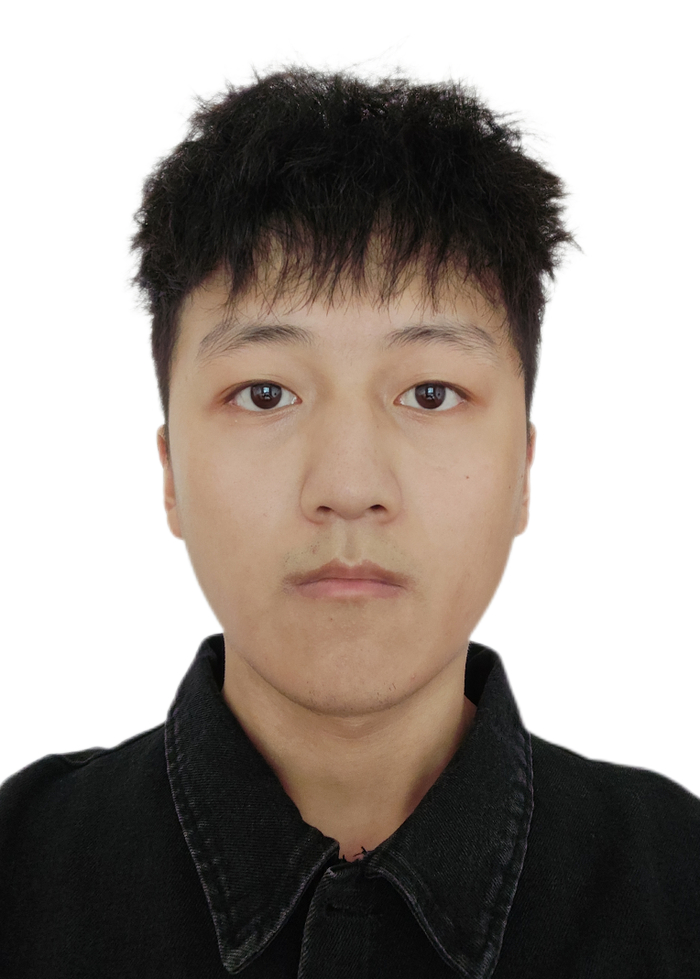}}]{Qianhao Gao}
	received the B.S. degree in communication engineering from the School of Internet of Things Engineering, Jiangnan University, Wuxi, China, in 2023. He is currently pursuing the Ph.D. degree in communications and information systems with College of Electronic and Information Engineering, Nanjing University of Aeronautics and Astronautics. His current research direction is related to spectrum mapping.
\end{IEEEbiography}

\vfill

\end{document}